\documentclass[pdflatex,sn-mathphys]{sn-jnl}

\usepackage{amsfonts,amsmath,amssymb}
\usepackage[T1]{fontenc}
\usepackage{bm}

\usepackage{subcaption}
\usepackage{graphicx}

\jyear{2021}%

\theoremstyle{thmstyleone}%
\theoremstyle{thmstyletwo}%

\theoremstyle{thmstylethree}%

\begin{document}

\title[Crystal-based pair production for a lepton collider positron source]{Crystal-based pair production for a lepton collider positron source}


\author[1]{\fnm{L.} \sur{Bandiera}}
\author[2,3]{\fnm{L.} \sur{Bomben}}
\author[1]{\fnm{R.} \sur{Camattari}}
\author[4]{\fnm{G.} \sur{Cavoto}}
\author*[5]{\fnm{I.} \sur{Chaikovska}}\email{iryna.chaikovska@ijclab.in2p3.fr}
\author[5]{\fnm{R.} \sur{Chehab}}
\author[6,7]{\fnm{D.} \sur{De Salvador}}
\author[1,8]{\fnm{V.} \sur{Guidi}}
\author[9]{\fnm{V.} \sur{Haurylavets}} 
\author[2,3]{\fnm{E.} \sur{Lutsenko}}
\author[2,3]{\fnm{V.} \sur{Mascagna}}
\author[1]{\fnm{A.} \sur{Mazzolari}}
\author[2,3]{\fnm{M.} \sur{Prest}}
\author[1]{\fnm{M.} \sur{Romagnoni}}
\author[2,3]{\fnm{F.} \sur{Ronchetti}}
\author[6,7]{\fnm{F.} \sur{Sgarbossa}}
\author[1,8]{\fnm{M.} \sur{Soldani}}
\author[1]{\fnm{A.} \sur{Sytov}}
\author[1,8]{\fnm{M.} \sur{Tamisari}}
\author[9]{\fnm{V.} \sur{Tikhomirov}} 
\author[3]{\fnm{E.} \sur{Vallazza}}



\affil*[1]{\orgname{INFN Ferrara}, \orgaddress{\street{via Saragat 1}, \city{Ferrara}, \postcode{44122}, \country{Italy}}}

\affil[2]{\orgdiv{Dipartimento di Scienza e Alta Tecnologia}, \orgname{Universit\`a degli Studi dell'Insubria}, \orgaddress{\street{via Valleggio 11}, \city{Como}, \postcode{22100}, \country{Italy}}}

\affil[3]{\orgdiv{Istituto Nazionale di Fisica Nucleare }, \orgname{Sezione di Milano Bicocca}, \orgaddress{ \street{Piazza della Scienza 3}, \city{Milan}, \postcode{20126}, \country{Italy}}}

\affil[4]{\orgdiv{Dipartimento di Fisica}, \orgname{Sapienza Univ. Roma and INFN Roma}, \orgaddress{\street{Piazzale A.Moro, 2}, \city{Roma}, \postcode{00185}, \country{Italy}}}

\affil[5]{\orgdiv{Université Paris-Saclay}, \orgname{CNRS/IN2P3, IJCLab}, \orgaddress{\city{Orsay}, \postcode{91405}, \country{France}}}

\affil[6]{\orgdiv{Dipartimento di Fisica e Astronomia}, \orgname{Universit\`a degli Studi di Padova}, \orgaddress{\street{Via Francesco Marzolo, 8}, \city{Padua}, \postcode{35121}, \country{Italy}}}

\affil[7]{\orgdiv{Istituto Nazionale di Fisica Nucleare}, \orgname{Laboratori Nazionali di Legnaro}, \orgaddress{\street{Viale dell'Università, 2}, \city{Legnaro}, \postcode{35020}, \country{Italy}}}

\affil[8]{\orgdiv{Dipartimento di Fisica e Scienze della Terra}, \orgname{Università degli Studi di Ferrara}, \orgaddress{\street{via Saragat 1}, \city{Ferrara}, \postcode{44122}, \country{Italy}}}

\affil[9]{\orgdiv{Institute For Nuclear Problems}, \orgname{Belarusian State University}, \orgaddress{\street{Bobruiskaya 11}, \city{Minsk}, \postcode{220030}, \country{Belarus}}}


\abstract{An intense positron sources is a demanding element in the design of future lepton colliders. A crystal-based hybrid positron source could be an alternative to a more conventional scheme based on the electron conversion into positron in a thick amorphous target. The conceptual  idea of the hybrid source is to have  two separate objects, a  photon  radiator and  a photon-to-positron converter target.  In such a scheme an electron beam  crosses a thin axially oriented crystal with the emission of a channeling radiation, characterized by a considerably larger amount of photons if compared to  Bremsstrahlung.  The net result is an increase in the number of produced positrons at the converter target.  
In this paper we present the results  of  a  beam test  conducted  at  the DESY  TB 21  with  5.6~GeV electron beam and a crystalline tungsten radiator. Experimental data clearly highlight an increased production of photons and they are critically compared with the outcomes of novel method to simulate the number of radiated photons, showing a very good agreement.
Strong of this, the developed simulation tool has been exploited to design a simple scheme for a positron source based on oriented crystal, demonstrating the advantages in terms of reduction of both deposited energy and the peak energy deposition density if compared to conventional sources. The presented work opens the way for a realistic and detailed design of a hybrid crystal-based positron source for future lepton colliders.}




\maketitle

\section{Introduction}\label{sec1}

Linear and circular colliders are nowadays  the most advanced instruments  to study fundamental particle physics.
Both the detailed study of the Standard Model (SM) of the fundamental interactions and the search for its possible extension (BSM) require future  colliders that could be the successors of the present CERN Large Hadron Collider.  

Electron-positron ($e^+ e^-$)  colliders offer  a very clean collision environment compared to hadron-hadron colliders which, however, could reach more easily a higher center of mass energy \cite{FCC_CDR_2,Chaikovska:2019ztn}.  High precision measurement of the SM processes might therefore  prefer a $e^+ e^-$ collider while  a direct observation of potential BSM particles requires higher colliding beam energies. A muon collider, despite the technical challenges to build it,  might accelerate beams to  multi-TeV energies with a relatively clean environment.


For all these future accelerators, however,  luminosity is the key factor to reach the desired precision.
For lepton colliders - $e^+ e^-$ and $\mu^+ \mu^-$ with muons produced by positrons \cite{Alesini:2019tlf,Amapane:2019oog,Amapane:2021npr,cesarini:2021} -  the availability of intense and small emittance electron and positron  beams is a crucial element to reach the required  luminosity. 
In particular, a positron source  represents a major challenge. 

\begin{figure}[h]
\centering
\includegraphics[width=\textwidth]{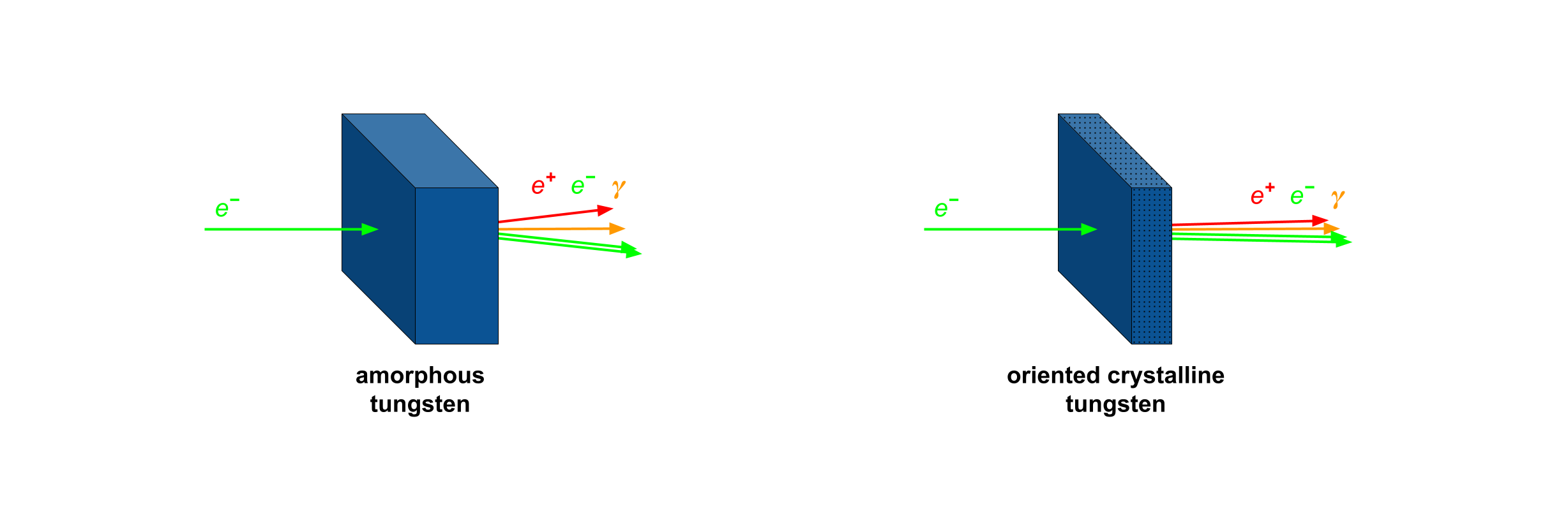}
\caption{Single-target schemes for the positron-source: conventional one with an amorphous tungsten target (that with no alignment with the impinging beam)  (left) and compact, crystal-based  and oriented target (right). Given its relative compactness, the latter configuration allows to attain lower positron emittance.}
\label{fig:sources_oneStage}
\end{figure}

For many years  a  conventional  way to realize a positron source consisted in using a  target with a high atomic number $Z$(as tungsten) hit by a high-energy primary electron beam. As shown in Figure \ref{fig:sources_oneStage} left, photons  are produced by Bremsstrahlung within the target  and are then converted in $e^+ e^-$  pairs.

Incident electrons  with high energy  and rather thick targets are needed to reach a large positron yield. The ideal target thickness should correspond to the maximum of the transition curve, which represents the secondary  particles yield as function of the target thickness. The SLAC linear collider, SLC, used  such a conventional source with a 30~GeV incident electron beam and  a 6 radiation length  ($X_0$) thick target. As observed for the SLC target,  important  heat load and high density of  energy  deposited in the target represented  a crucial problem~\cite{sheppard2003conventional}.


A high rate of $e^+ e^-$  pairs produced in the target depends on the generated  photon yield.  Besides those based on the Bremsstrahlung process, others photon sources are considered for the positron sources necessary for $e^+ e^-$ colliders, such as undulator  radiation and  Compton backscattering radiation, which may provide polarized photons~\cite{STRAKHOVENKO2005320, chaikovska2012polarized}. New kinds of photon sources have  also been studied for unpolarized positrons, for example the planar undulator  radiation for the unpolarized version of the TESLA project~\cite{balewski1991status, holtkamp1995status} and the channeling radiation. Here, we focus our attention on the possible development of a positron source for future colliders based on channeling and its main advantages.

\subsection{Channeling radiation in crystals}
Radiation processes in oriented crystals opened a vast and promising field for the photon production.


Indeed, light particles as $e^+$ or  $e^-$   penetrating a crystal  with a small angle with respect to the atomic rows (crystal axes) or lattice planes are subject to very strong electric fields generated by the  atoms of the crystal. They, therefore,  emit  a radiation  that is defined as  {\it coherent} since it is the result of the coherent interaction of the single $e^+$ or  $e^-$  with all the atoms in the crystal~\cite{KUMAKHOV197617}. This radiation is particularly interesting  when the particles are penetrating the crystal along one of its main symmetry axes. The  particles with this orientation  are  said to be channeled or quasi-channeled.  The most intense radiation  arises from particles in the channeling condition, i.e. with angles smaller than the critical angle $\Psi$,
\begin{equation}
    \Psi = \sqrt{\frac{2U_0}{E}},
\end{equation}
where $U_0$ is the atomic potential well depth and $E$ the particle energy. For instance, $\Psi$ is about 0.5~mrad  in a  tungsten crystal kept at normal temperature, oriented along its <111> axis and hit  by a 10~GeV electron beam parallel to this axis. 

Such radiation has a larger power than the radiation produced from the same crystal but  with a random orientation with respect to the particle direction. In particular, at high enough incident particle energy, the energy radiated in channeling orientation is significantly enhanced if compared to standard Bethe-Heitler Bremsstrahlung.  This enhancement depends on the  type of the crystal, the axis of symmetry and on the particle ($e^-$) energy. As an example it appears at an energy  larger than 0.7~GeV for  tungsten,  and even larger  for Si and Ge (1.3 and 1.9~GeV, respectively \cite{Baier_Katkov_Strakhovenko}). 

It must be noted that for GeV energy photons the  $e^+ e^-$  pairs production in  crystals is well described by the Bethe-Heitler mechanism. However, for higher photon energies ($E$ > 20~GeV, for W crystal, $E$ > 100~GeV for Ge,  with both crystals  kept at normal temperature)  the on-set of another effect - the  pair production  in strong fields -  appears,  leading to very large enhancements. As an example, an order of magnitude larger  pair production compared to the  Bethe-Heitler mechanism has been observed at CERN with 150~GeV photons impinging on a Ge crystal cooled at 100~K \cite{PhysRevLett.53.2371}. In the application for positron sources  described in this paper the most common value for the incident electrons energy is of few GeV, therefore well below the threshold for pair production in strong fields. 

In any case, for the energy range of interest an effective radiation length can be introduced to describe the channeling radiation, representing the  mean distance over which an electron loses all but $\frac{1}{e}$ of its initial energy by channeling radiation. This distance is in fact much shorter than the standard $X_0$, which describes  the typical length scale of the Bremsstrahlung production.
In fact  an electromagnetic  shower development in a crystal deserve a specific description \cite{BAIER1995147, BAIER1999403, Bandiera18}.

At the energies of interest, channeling radiation is composed essentially of soft photons \cite{ARTRU1994443}. These photons generate mainly soft $e^+e^-$ pairs (from some MeV to some tens of MeV kinetic energy). It must be underlined that the soft positrons are the most useful for the positron sources as they are easily captured by existing capture systems~\cite{Chehab:197428} for particle accelerators.

\subsection{Crystal-based positron sources}


A positron source driven by channeling radiation (Figure \ref{fig:sources_oneStage} right) was first proposed in 1989~\cite{chehab1989study}. Photons production in Ge and Si crystals and their subsequent conversion into pairs in a $W$ amorphous target were studied. For a 20~GeV $e^-$ beam impinging on a 1~cm thick Ge crystal the photon yield was above 20~$\gamma$/$e^-$. A proof of principle experiment at the Orsay Linac (2~GeV) confirmed the relevance of this way to produce photons \cite{ARTRU1996246}.

The investigations on this kind of source with the associated simulations rely on theoretical works developed at the Budker Institute (Baier-Katkov-Strakhovenko) and by a theoretician from LPT-Orsay and IPN-Lyon. Simulation programs were developed to describe more carefully the processes.  These simulations were central to interpret the data collected by the experiments~\cite{BAIER1995147, BAIER1999403}.
The simulation used to describe the channeling radiation relies on the Baier-Katkov formula for radiation in non-uniform fields~\cite{baier1968processes, Baier_Katkov_Strakhovenko0} that considers the complete electron trajectory in the crystal. Another approach - instead of taking into account  the whole trajectory - led to the  {\it FOT}  code developed by X.~Artru, where the integration of the Baier-Katkov probability  involves only parts of the trajectory  avoiding excessive  computing time \cite{XAVIER1990278}. Other simulation programs like {\it SGC } (Shower Generation in Crystal) developed by V.M.~Strakhovenko are taking into account the electromagnetic interaction in axially oriented crystals. They are all able to describe also the shower development in amorphous targets~\cite{BAIER1999403}. 

Radiation and pair production methods have been developing in the Institute for Nuclear Problems (INP) in Minsk since 90-th \cite{Tikhomirov01, Tikhomirov02, Tikhomirov03, Tikhomirov04} and, for the last decade, jointly with a Ferrara group \cite{Bandiera19}, which developed its wide experimental verification. During the last decade a lot of different orientational effects in crystals have been simulated by the collaboration of the Ferrara and INP groups. Their results are in fair comparison with experiments \cite{Bandiera19, Bandiera21, Bandiera22, Bandiera24, Bandiera25, Sytov2, Mazzolari1, Wistisen1, Sytov3}.

An experiment at CERN (WA~103) was performed in 2000-2001 with the aim of finding the actual positron yield provided by the sources using channeling. This experiment used thick tungsten crystals (4 and 8~mm thick), where photon generation and pair production occurred in the same target. The experimental results were compared to the simulations and allowed their validation \cite{CHEHAB200241, ARTRU2005762, ARTRU2003243}. 
Positron source with the crystal converter was also investigated experimentally at KEK~\cite{PhysRevE.67.016502, PhysRevSTAB.10.073501}.
Though, as demonstrated by WA~103 experiment, the positron yield obtained with thick crystals corresponded to the requirements for a linear collider positron source~\cite{ ARTRU2005762}, the problem of the heat load in crystals remained outstanding. 

Using thick crystals may, in fact, lead to an important energy deposition in the crystal itself associated to the shower development. Besides the problem of the survival of the crystal for high deposited power, this has the consequence of reducing the channeling potentials in the crystal due to the increased thermal vibrations. Eventually, this results into a decrease of the expected enhancement. Simulations using Baier’s formula for the potential  \cite{Baier_Katkov_Strakhovenko} allowed estimation of the possible losses in positron yields \cite{Artru:1997fy}. 

\begin{figure}
\centering
\includegraphics[width=1.\linewidth]{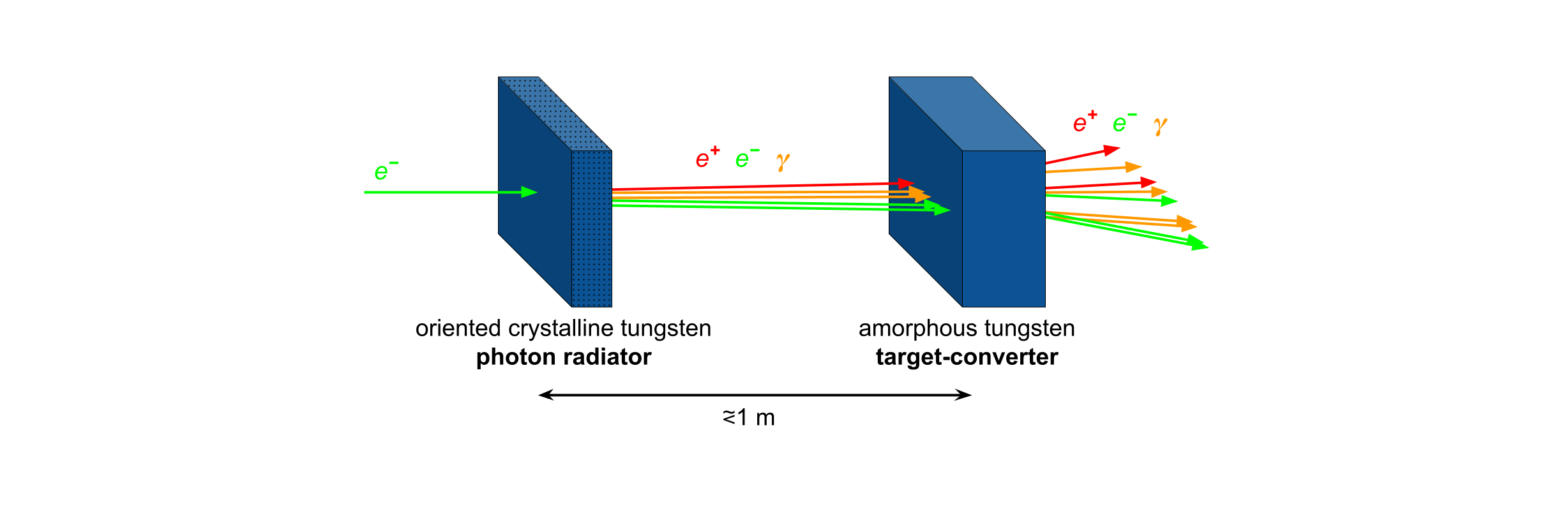}
\includegraphics[width=1.\linewidth]{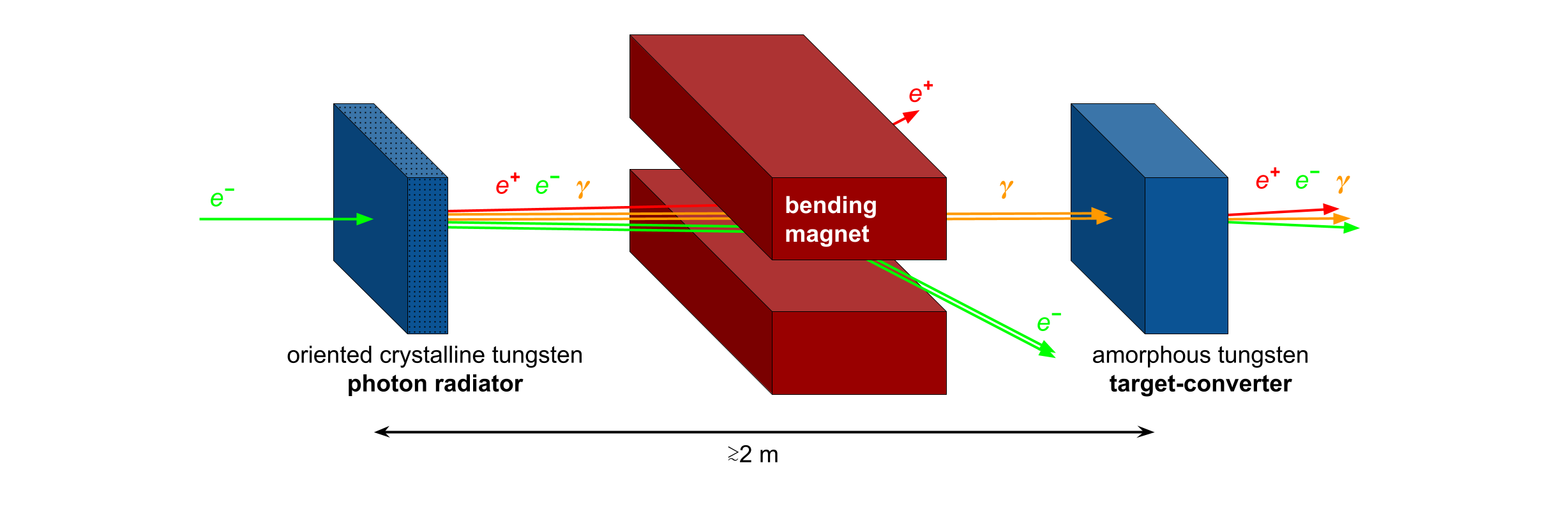}
\caption{Hybrid schemes for the positron source. Simple two-stage version, which features a crystalline photon radiator followed by an amorphous converter (top), and optimized version, in which a bending magnet is placed downstream with respect to the photon converter to redirect all the $e^+e^-$ generated in the first stage away (bottom).}
\label{fig:sources_twoStage}
\end{figure}

Henceforth the positron source using channeling is better conceived as a compound or hybrid target with two elements, a crystal  with the function of radiator followed downstream by an amorphous target acting as converter of photons into $e^+e^-$ pairs \cite{Artru:2008zz}, as sketched in Figure \ref{fig:sources_twoStage}. Studies and experiments, thus, enhanced the interest on such sources~\cite{Dadoun:1248436, chehab2011posipol,artru2011positron, SATOH20053, SUWADA2006142,dadoun2012event, UESUGI201417, ARTRU201560, CHAIKOVSKA201758}.

Rather thin crystals with a thickness of 1-2 mm in the considered energy range may provide a high rate of photons. In this way the heat load in the crystal itself is limited. The thick converters (few $X_0$) are subject to a much more important heat load and its level and the power density must be seriously studied. 

The energy deposition in the converter is far from being homogeneous, leading to possible thermal stresses. The stresses are particularly strong on the exit part of the converter, where the electromagnetic shower is at its maximum and presents a peak of the deposited  power density. The breakdown of the SLC target due to the thermal stresses led to the systematic determination of the Peak Energy Deposition Density (PEDD). The studies undertaken at LANL and LLNL~\cite{maloy2001slc, stein2001thermal} concerning the SLC target showed that the PEDD  of 35~J/g might not be exceeded. This is now considered as a maximum tolerated value of the PEDD in the tungsten targets for a positron source design.
These  problems have been reviewed in~\cite{PhysRevSTAB.6.091003}. Investigations on hybrid positron sources have been also carried out by different  groups~\cite{azadegan2013positron}. In order to more accurately design a hybrid positron source further investigations are needed. 
Improvement in hybrid sources regarding the thermal stresses led to the choice of a granular converter, where the shock waves associated to the stresses can be considerably reduced~\cite{cheng2012positron, Chaikovska:IPAC17-WEPIK002}.

In this paper we present the experimental investigation of the channeling radiation produced in an oriented 2 mm tungsten crystal by the passage of 5.6~GeV electrons. In particular, the total energy lost by each electron crossing the crystal in the form of radiation has been measured. At the same time a fraction of this radiation is converted into $e^+e^-$ pairs by a thin copper converter and the $e^+e^-$ pairs are directly detected and counted. The main goal of this experimental test was to validate our improved simulation toolkit that includes a better description of the number of photons composing the channeling radiation. The results of a simulation of the processes of radiation production under channeling condition are reported and validated with data along with a full Geant4 simulation of the experimental apparatus. In particular, the experimental signal of the photon conversion system to \textit{count} the photons was well reproduced by full simulations. Strong of this, we exploit simulation to investigate the performance of a hybrid source and compare it with a conventional source. The energy case selected for this study was the one of interest for future colliders included in the European Strategy for High-Energy Physics.%

\section{Crystal characterization at the DESY T21 beamline}
\label{section:desy}


A study of the radiation emitted by a high-quality tungsten crystal was performed using the DESY beamtest facility T21 \cite{2019_DESY} where a 5.6 GeV electron beam was available.

\subsection{Tungsten crystalline target  }

\begin{figure}[h!]
\begin{center}
\includegraphics[width=0.7\textwidth]{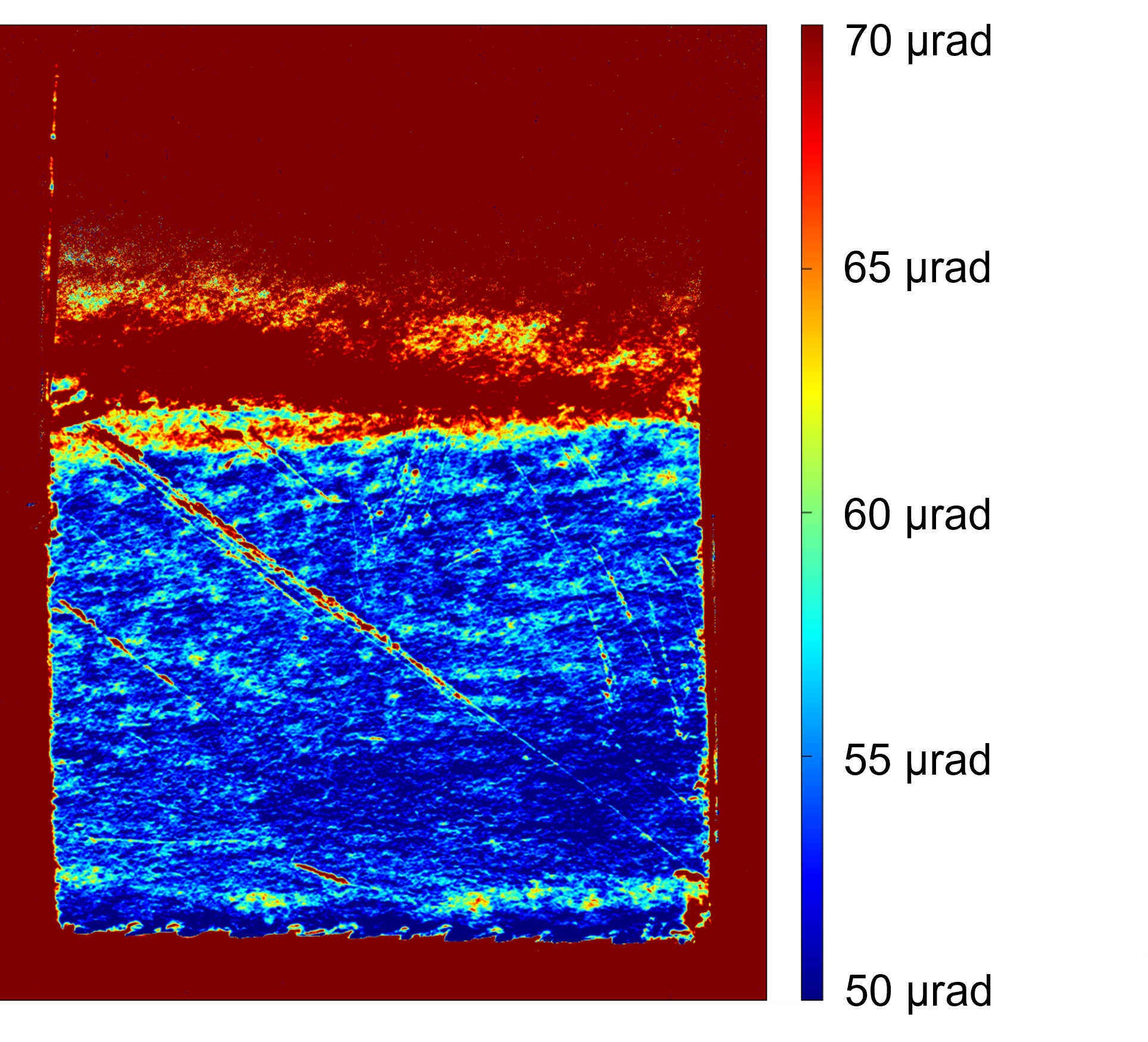}
\caption{Imaging of the sample mosaicity measured at BM05 beamline of ESRF (Grenoble, France). Color indicates the mosaicity of the sample.}
\label{fig:Wxrays}
\end{center}
\end{figure}

The tested W was $2.25 \pm 0.05$~mm thick (equivalent to $0.65 X_0$), with a $\sim 7 \times 7$~mm$^2$ square transverse section. This crystal was manufactured by the Laboratory of Materials Science (LMS) in the Institute of Solid State Physics of the Russian Academy of Science (ISSP RAS). The crystal was installed with its $\langle 100 \rangle$ axial direction oriented along the beam axis. For this axial orientation $\Psi$ = 0.52 mrad. The lattice quality of the sample was estimated via $X$-ray diffraction, in particular the mosaicity (i.e. the mean angular spread of crystallites within the sample) was assessed. Indeed, this is a critical quantity affecting the effective performance given the strong angular dependence of the axial effect. The measurements were performed at the synchrotron facility ESRF (beamline BM05). The facility allowed both high spatial ($\approx 5 \mu m$) and angular ($\approx 1 \mu rad$) resolution thanks to the high intensity and energy of the diffracting photons (20 keV). Analysis of x-rays diffraction intensity recorded as the crystal rotated in and out of Bragg diffraction condition allowed to define mosaicity on each position of the sample surface (as shown in Figure \ref{fig:Wxrays}) . The results show a remarkably low mosaicity, with most of the crystal showing values $\leq 60~\mu\mathrm{rad}$. Region of largest mosaicity are still below $150~\mu rad$, and are found in correspondence to small surface scratches, thus are most probably limited to surface of the sample. Ultimately, for the entire sample the mosaicity observed is well below the critical angle, hence the crystal is well suited for investigation of axial effects.

\subsection{Experimental apparatus }
Electron beams are obtained from the DESY II synchrotron by a double conversion: firstly, Bremsstrahlung photons are generated by a carbon fiber target positioned in the synchrotron beam orbit; then, these photons hit a secondary target generating $e^+e^-$ pairs. From the latter, electrons with the chosen momentum are selected with a dipole magnet. The resulting electron beam comes in two $20$--$40$~ms long (energy-dependent) bunches for each $160$~ms long DESY II cycle. The DAQ (Data AcQuisition) exploited in this beamtest was set to work on $\sim 10$~s long cycles with the first $\sim 5$~s for the actual acquisiton, the remaining time being exploited to write the data on disk. A $5.6$~GeV beam was used for the studies described here, with a measured rate of $\sim 2$--$5 \times 10^3$~particles/s at the upstream tracking detector. This rate allows a single particle tracking with our apparatus.

\begin{figure}[htbp]
\centering
\includegraphics[width=\textwidth]{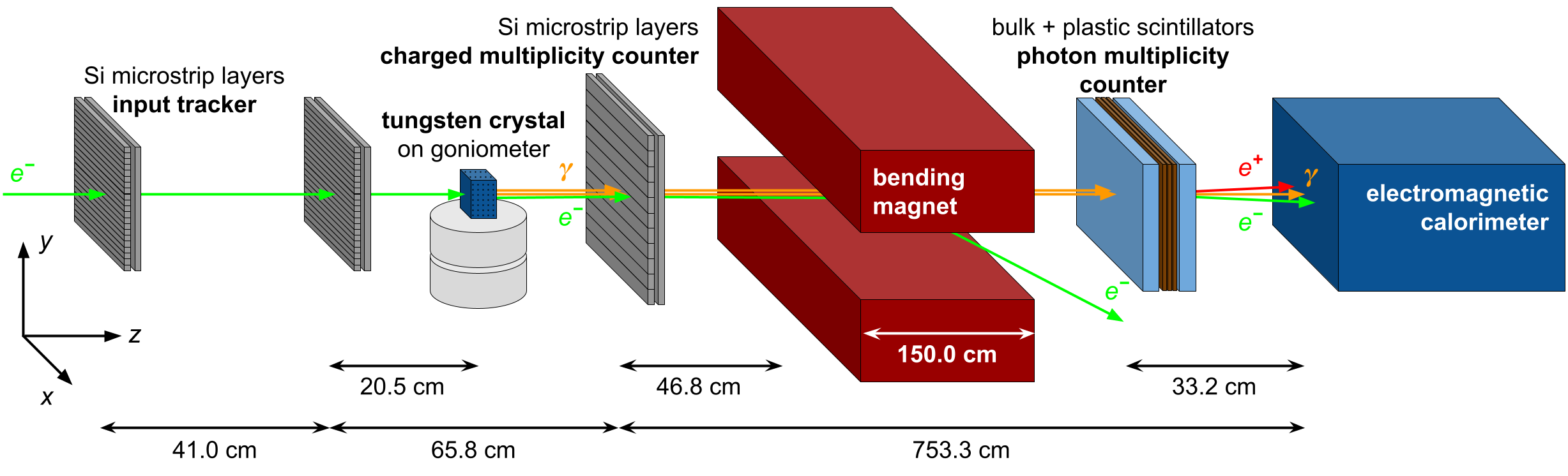}
\caption{Scheme of the DESY experimental apparatus.}
\label{fig:DESYexp}
\end{figure}

In Figure \ref{fig:DESYexp}, the experimental apparatus installed on  T21 lines for the crystal studies is schematically outlined. The apparatus was composed of two sections --- with a magnet in between the two.

In the first section two silicon double-sided microstrip detectors with a $2\times2$~cm$^2$ area and a spatial resolution of slightly more than $10~\mu$m \cite{2013_Lietti} at a relative distance of $41$~cm along $z$ were installed upstream of the crystalline targets. They were used to reconstruct the trajectories of the input particles, and hence their incident angle with respect to the crystal surface, and to monitor the transverse $x$--$y$ size of the incoming beam and its angular divergence in both the $x$--$z$ and $y$--$z$ planes. In order to observe the channeling effect, the beam angular divergence should be at most of the order of the channeling critical angle $\Psi$.
The beam size was much larger than the sample size and a flat distribution at the target $z$ position. The distribution of the electron track angles was found to be approximately Gaussian with a standard deviation of $780$ ($720$)~$\mu$rad in the $x$--$z$ ($y$--$z$) plane.
 
The crystalline target was mounted on a high-precision goniometer equipped with two linear stages (vertical and horizontal) to move the crystal under test into the beam path and two rotational stages to orient its lattice planes or axes with respect to the average beam direction \cite{2021_Soldani}. The crystal was mounted on a light plastic support that was built to guarantee a pre-alignment of the order of 1 mrad at the installation on the rotational stages.

Photons and secondary $e^+e^-$ pairs were produced in the crystal by the passage of the $5.6$~GeV electrons. Downstream the goniometer (at a $45.3$~cm distance) a couple of silicon $10 \times 10$~cm$^2$ area layers with a double-hit resolving power of about $1$~mm were used as charged track multiplicity counter. This detector was used to identify the condition of alignment with the target lattice planes or axes, which has the distinct feature of a secondary pair production enhancement with respect to random orientation. This alignment conditions were usually obtained by taking data while the goniometer angular positions are continuously changed ({\it angular scan}). 

A dipole magnet was operated to generate a $1.339$~T vertical (i.e., along $y$) magnetic field that was uniform over a $150$~cm long span along $z$. When the dipole magnet was powered, the primary electron beam and all the smaller energy secondary charged particles generated inside the crystal were swept away from the $z$ axis and thus separated from the emitted photons, which then were propagating down to the second section of the apparatus. This second section featured a series of scintillators, silicon trackers and - downstream to the whole apparatus - a crystal electromagnetic (e.m.) calorimeter.
 
A $10 \times 10$~cm$^2$, $4$~cm thick scintillating detector was installed right after the magnet to further suppress the possible background induced by charged particles entering the second section, mostly from photon conversion in the $\sim 7$~m of air and in other bulk layers across the photon path, within and at the exit of the magnet. Indeed, events with a signal in this scintillator were vetoed in the offline data analysis.

\begin{figure}[htbp]
\centering
\includegraphics[width=0.55\textwidth]{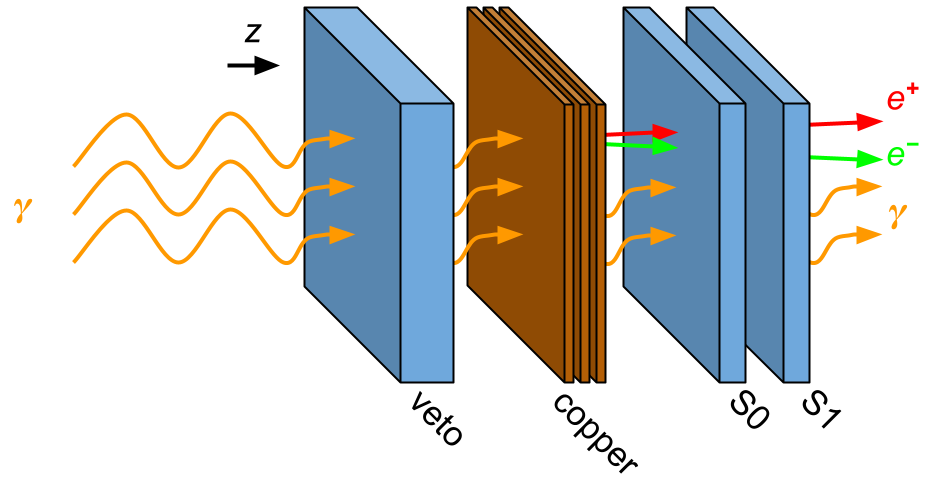}
\caption{Scheme of the active photon conversion system. The veto scintillator is also sketched.}
\label{fig:preshSketch}
\end{figure}
 
An active photon converter was used to estimate the average number of photons produced per electron interacting in the crystalline target at different angular orientations. A sketch of this detector is shown in Figure \ref{fig:preshSketch}. The copper layer enhances the probability of conversion into $e^+e^-$ pairs. An overall thickness of $2.7$~mm Cu ($0.2 X_0$) was chosen for these  layers. The number $N_{ee}$ of $e^+ e^-$ pairs generated inside the copper layer is then measured by a pair of $10 \times 10$~cm$^2$, $1$~cm thick scintillating detectors (S0 and S1 in Figure \ref{fig:preshSketch}). 

Eventually, the full energy of the radiation exiting the crystalline target, $E_{loss}$, is absorbed and measured by the e.m. calorimeter, only a small fraction of it being lost in the upstream material. The calorimeter consists of a $3 \times 3$ matrix of tapered, $20.5 X_0$ long BGO crystals with a square front transverse section of $2.1 \times 2.1$~cm$^2$ and a PMT-based readout --- each crystal was coupled to a Photonis XP1912 PMT \cite{abbiendi2021study}.

A calibration of the calorimeter response was performed by removing the target and switching off the dipole magnet. In this configuration, the electron beam directly impinged on the BGO calorimeter. Data were collected at various beam energies between $2$ and $5.6$~GeV. A full simulation was performed with the Geant4 toolkit \cite{2003_geant} at all the energy points, from which the value of energy deposited in the calorimeter corresponding to each point was obtained. The signal equalization of the nine BGO crystals and the calibration of the total signal, performed with the aforementioned deposited energy values, led to a resolution on $E_{loss}$ of $\sim 800$~MeV at $5.6$~GeV \footnote{Better performance in terms of resolution could be attained by exploiting muons (unavailable in this experimental facility) for the channel equalization --- details can be found, e.g., in \cite{abbiendi2021study}}.

The same beam configuration was exploited, at $5.6$~GeV, to characterize the response of the active photon converter to the passage of a single electron (whose deposited energy is that of a Minimum Ionising Particle, or MIP), so that a relation between the energy deposited in it and $N_{ee}$ could be established. Again, a comparison with the full simulation was made, and an average energy deposit of $1.67 \pm 0.01$~MeV was estimated for a single MIP in both scintillators.

\begin{figure}[h]
\centering
\includegraphics[width=\textwidth]{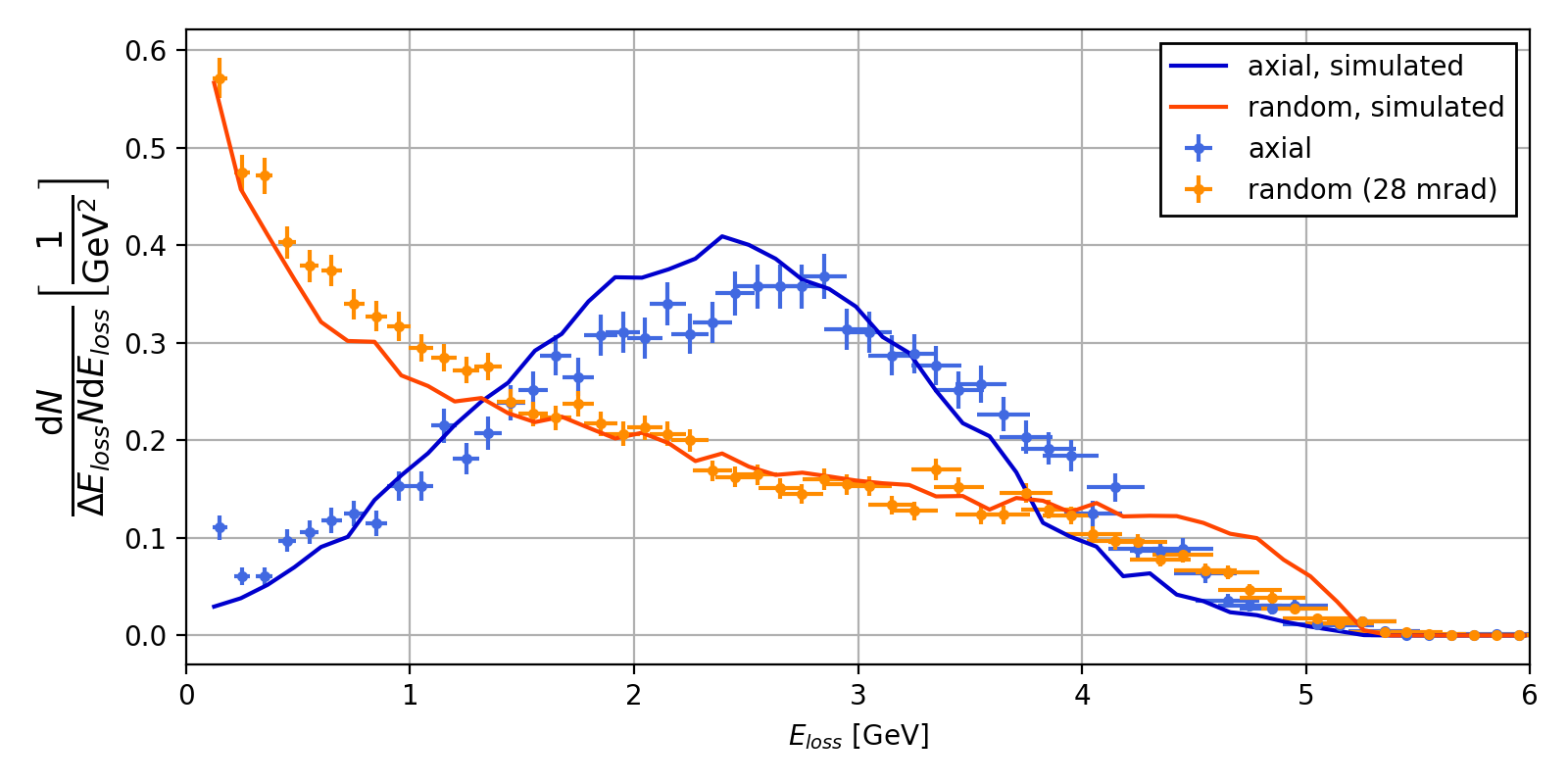}
\caption{Spectrum of $E_{loss}$ in the axial (blue) and random (orange) angular configurations. Both the experimental data collected at DESY and the corresponding simulated distributions are shown.}
\label{fig:Eloss}
\end{figure}

\subsection{Results and comparison with simulation }

The $W$ crystal was probed with the particle beam for about $100$ hours, at various orientations of the crystal with respect to the beam.

Figure \ref{fig:Eloss} shows the spectra of $E_{loss}$ obtained in two very different angular configurations, i.e., on axis (blue dots) and at $28$~mrad away from the axis (orange dots). This is in fact an orientation sufficiently far from the crystal main axes to consider the atom in the crystal as randomly distributed.  The latter (orange) distribution  clearly shows the standard Bremsstrahlung spectrum, which populates the whole phase space continuously and decreases as $E_{loss}$ grows. On the other hand, the axial curve (blue) shows a significant suppression of the low-energy component of the spectrum with respect to the random orientation condition and a strong enhancement at higher energies (i.e., at $\gtrsim 1.4$~GeV) reaching  a maximum at $\sim 2.5$~GeV. The errors on the ordinates are statistical, whereas those on the abscissas are dominated by the calorimeter energy resolution.

\begin{figure}[h]
\centering
\includegraphics[width=\textwidth]{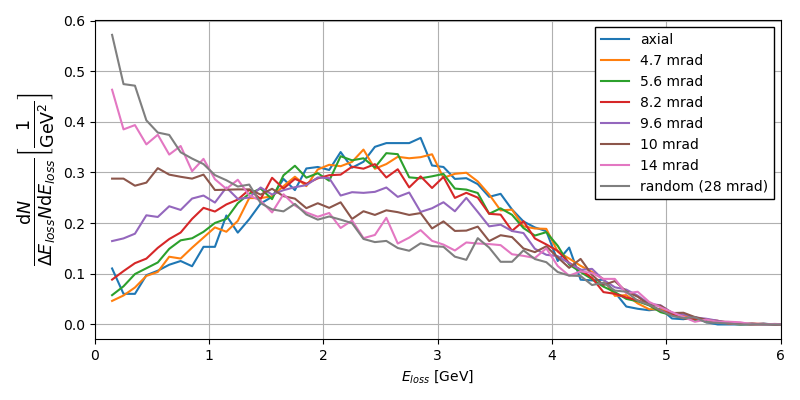}
\caption{Comparison between experimental spectra of $E_{loss}$ at various angular configurations.}
\label{fig:AxAm}
\end{figure}

The $E_{loss}$ spectrum transition from the axial configuration to the random one was also studied by performing measurements at different angles. Figure \ref{fig:AxAm} shows all the resulting spectra. It is clear that the strength of the coherent effects decreases only slightly up to $\sim 8$~mrad from the axis.

Since the $W$ crystal thickness is comparable with its radiation length, a detailed simulations of the e.m. shower developing in the crystal, both when oriented and randomly aligned is necessary. E.m. shower in an oriented crystal can be simulated by using the Baier-Katkov quasiclassical method \cite{Baier_Katkov_Strakhovenko0}.

This method includes a multidimensional numerical integral over a simulated particle trajectory and over the angles of the radiation emission direction. It takes into account quantum recoil of $e^+$ and  $e^+$ in the photon emission but is based on a classical trajectory. 

Both radiation and pair production processes can be simulated by means of the Baier-Katkov method. The main difference is a necessity of detailed trajectory simulations for radiation, while for pair production one may use an energy-dependent cross-section. The simulations of classical trajectories of $e^+$ and $e^+$ are valid down to 100-200 MeV, while below this energy threshold standard Bremsstrahlung cross-sections should be used.

All these features have been included in the simulation code \cite{Bandiera19, Tikhomirov1, Baryshevsky1}. In particular, to describe reliably the continuous process of soft photon formation in the field of multiple atomic strings, the Baier-Katkov formula integration has been extended to the trajectory parts of several dozens of micrometers. Being scattered up to the considerable angles, soft radiated photons will form at rather long coherence lengths crossing multiple atomic strings. At the same time, since incoherent scattering is quite important in high-Z crystals, the integration over each single trajectory part was computed discretizing the path into much shorter intervals that separate the incoherent scattering points, in which the radiating particle velocity changes abruptly, hampering the application of the continuous function integration methods.
Since the W crystal is thick enough for allowing multiple photon emission by primary electrons and pair production by emitted photons, in order to simulate a realistic photon yield one should continuously trace the radiation of all the low-energy electrons and positrons. Lots of such electrons are in the energy range of a few hundred MeV and will mostly move far from the channeling conditions. That is why, in order to speed up the simulation, these events can be generated by adopting the well verified and optimized procedures from the GEANT4 toolkit \cite{2003_geant}.

The code used here with $W$ crystalline target, has been already validated for e.m. shower in a lead tungstate crystal exposed to 120 GeV/c electrons at CERN \cite{Bandiera18}. 

In order to simulate the full experimental setup shown in Figure~\ref{fig:DESYexp}, the Geant4 simulation toolkit \cite{2003_geant} has been applied for both axial and random cases. For axial orientation the output of simulations of e.m. shower in oriented crystal containing coordinates and momenta of both primary and secondary particles at the crystal exit has been used as an input of Geant4. For the random case Geant4 has been also applied for e.m. shower simulations in the crystal.

The output of Geant4 simulations includes the  deposited energy in the BGO calorimeter as well as in scintillators and this deposited energy is then considered to be proportional to the detector signal.

Figure \ref{fig:Eloss} displays the simulated curves (continuous lines), obtained with a beamtest full simulation, in comparison with the experimental results. The overall agreement between simulation and data is in general satisfactory. 

An investigation on the number of photons produced per interaction between the beam particles and the crystalline sample as a function of their relative orientation was also performed during the experiment. This was done using data from the active photon converter.  Indeed, the signal in its downstream stage  should be proportional to the number of charged particles originating in the Cu layer, and hence to $N_{ee}$. Although the pair production by photons is intrinsically stochastic and the selected copper layer had limited thickness ($0.2 X_0$ only) in order to minimize the contributions by low-energy particle stopping and secondary Bremsstrahlung emission (i.e., the start of an e.m. shower), this setup was expected to be highly sensitive to the variation in the distribution of the number of photons coming from the crystal, which is closely related to that of $N_{ee}$, at different lattice orientations. The idea was taken from a previous experiment \cite{CHEHAB200241} in which thick W crystals were tested in the Single-target scheme configuration (see Figure~\ref{fig:sources_oneStage}) and where a qualitative enhancement of the signal at the preshower was measured in case of axial alignment. The main novelty here is given by the possibility to reproduce the experimental results with our optimized simulation tool and thereby to eventually extrapolate the radiation spectrum and use it to design a realistic positron source.

\begin{figure}[htbp]
\centering
\includegraphics[width=\textwidth]{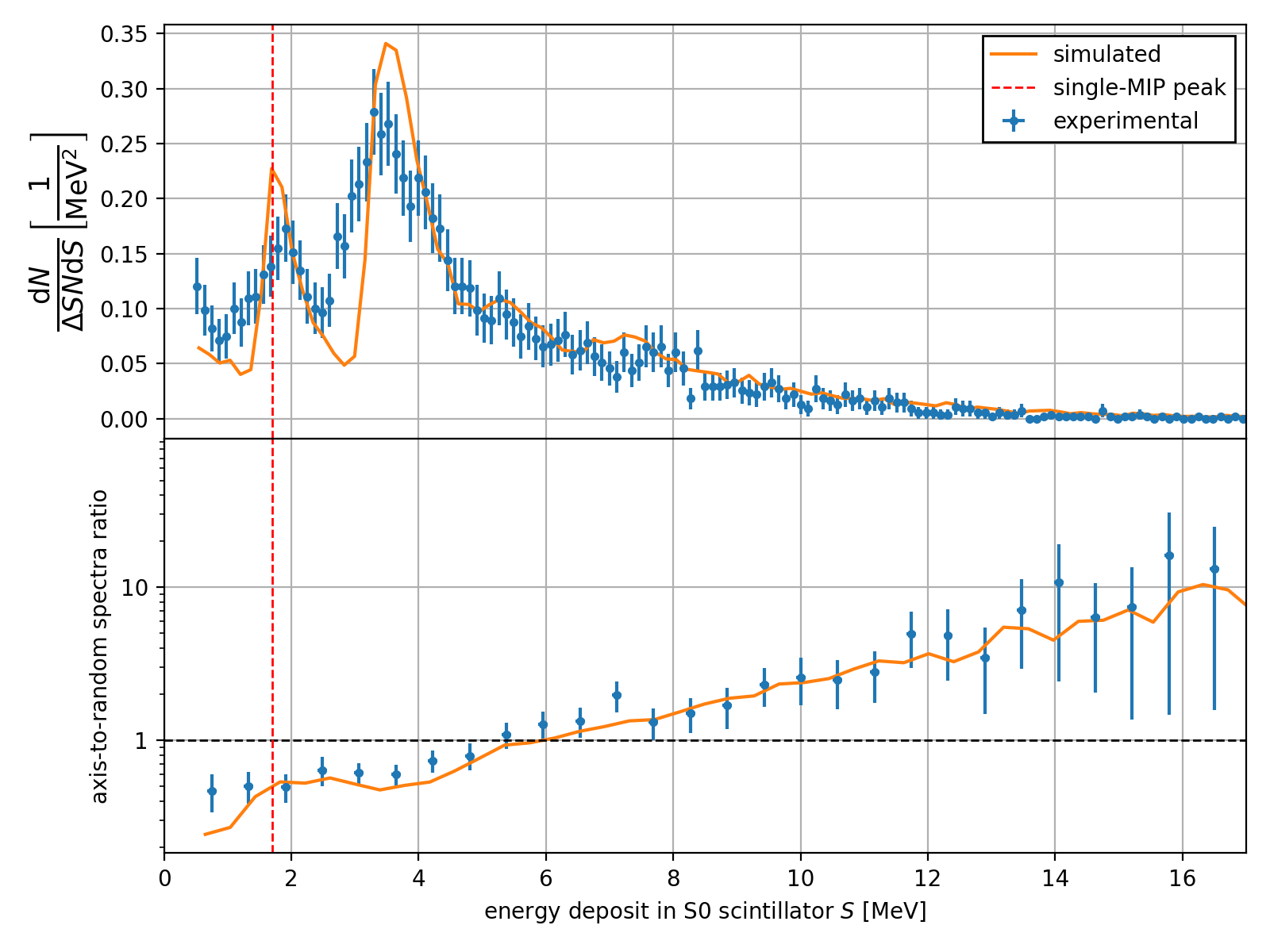}
\caption{Top: spectrum of the energy deposited in scintillator S0 (i.e., the upstream output scintillator of the active photon converter) when the crystalline sample is in random configuration. Bottom: ratio between the S0 signal spectra in axial and in random configuration. The experimental (simulated) distribution is shown in blue (orange), and the single-MIP signal peak value is indicated with a dashed red line.}
\label{fig:preshData}
\end{figure}

Figure \ref{fig:preshData} shows the results obtained with data from S0, i.e., the upstream scintillator in the active photon converter output stage (see Figure \ref{fig:preshSketch}). The results for the scintillator S1 are not presented since they lead to nearly identical outcome. As an example, the spectrum of energy deposited in the scintillator is shown for the random case (upper plot): the single-MIP (i.e., single-electron or -positron) peak is clearly visible and compatible with the reference value found in the single-track calibration data described above (dashed red line), and higher-multiplicity quantization can be observed up to $4$ MIPs (corresponding to $\sim 7$~MeV). The bottom plot in Figure \ref{fig:preshData} shows the ratio between the distributions obtained on axis and in random orientation. As expected, when in axial configuration, the fraction of events with less than $3$ MIPs is suppressed with respect to the random case, whereas the higher-multiplicity component is boosted up, thereby demonstrating an increase in the number of photons. Such boost is observed to become ten-fold when the energy deposit reaches a value of $\sim 16$~MeV, corresponding to $9$--$10$ MIPs and, hence, to $N_{ee} \sim 5$. We assume that reasonably the scintillator response stays linear up to several MIPs. Again, Figure \ref{fig:preshData} highlights the good agreement with the Geant4 simulation. It has to be noted that the latter doesn't reproduce the error contribution of the finite detector resolution, which makes the peaks in the top plot higher and narrower than the experimental ones.

The results shown in Figures \ref{fig:Eloss} and \ref{fig:preshData} demonstrate the very good compatibility between the experimental data and the outputs of the beamtest simulation. This proves the reliability of the software tools that have been used to simulate the e.m. interactions in oriented crystals. These tools could therefore be exploited for the studies presented in Section \ref{section:PositronSourceFullSim}.


\section{Conventional and hybrid schemes for positron production}
\label{section:PositronSourceFullSim}


As illustrated in the previous Section the developed simulation tool of coherent interactions leading to radiation emission in a crystal has been validated with data taken at the DESY TB facility with 5.6 GeV electrons interacting with a tungsten crystalline target, which is of great interest for intense positron sources for future colliders.
Therefore, to illustrate the advantages of a crystal-based  source two scenarios using the conventional and the  hybrid targets (see Figure~\ref{fig:sources_oneStage} and \ref{fig:sources_twoStage}) have  been studied. An  electron beam  of  6~GeV  energy has been here considered for a hypothetical positron source being such configuration intersting for the FCC-ee current design. The hybrid scheme involving two targets without magnet (Figure~\ref{fig:sources_twoStage} up) in between is studied due to potential applications at the circular colliders, where the beam power is considerably lower compared to linear colliders, where the use of the magnet to sweep away the charged particles is of greater importance (Figure~\ref{fig:sources_twoStage} bottom).

The first step is to optimize the crystal target for the hybrid scheme.
The simulations for radiation and pair production in an oriented crystal and amorphous case (random orientation) were performed with the simulation code \cite{Bandiera19, Tikhomirov1, Baryshevsky1} and Geant4 code~\cite{2003_geant} respectively. In both cases the targets are made of tungsten $W$ material. As previously stated, the $W$ crystals provide a deep atomic potential due to the high value of atomic number $Z$.
The orientation of the crystal is fixed to be along the <111> axis allowing for higher values of the atomic potential compared to the <100> axis described in previous chapter (and used for the DESY experiment).

\begin{figure}[h]
\centering
\includegraphics[width=0.5\textwidth]{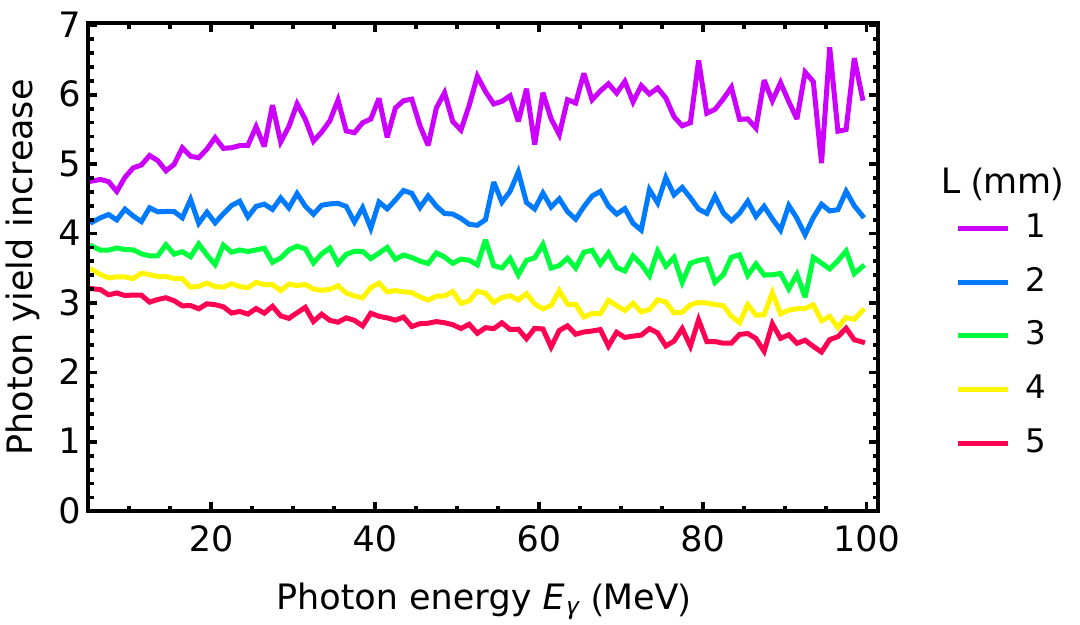}
\caption{Radiation enhancement in the tungsten (W) crystal aligned along <111> axes compared to the amorphous target having the same thickness (random orientation of the crystal). Electron beam energy is 6~GeV.}
\label{fig:crystalSIMU}
\end{figure}

Figure~\ref{fig:crystalSIMU} shows the simulation of the enhancement of the photon yield in the oriented crystal with respect to an amorphous target (that is the crystal with random orientation) having the same thickness, considering only photons with energy lower than 100~MeV. This energy range has been selected since these photons are responsible for the production of positron within the typical capture system acceptance~\cite{Chehab:197428}, i.e., few tens of MeV. For a 1~mm thick  $W$ crystal, the number of photons with energies lower than 100 MeV is 5.5 times higher compared to the photon production in the amorphous target. It can be also observed that the enhancement is higher for thinner crystals.
 
The number of photons produced for different crystal thicknesses is presented in Table~\ref{tab:photon_spectrum}. The results are given for two energy domains: full energy spectrum and low-energy part ($E$< 100~MeV) of the spectrum.
\begin{table}[]
\centering
\caption{Simulated number of photons produced by one incident 6~GeV electron for two tungsten (W) crystal orientations: crystal aligned along <111> axes and random orientation of the crystal. Low energy cutoff used in simulations is 5~MeV.}
\begin{tabular}{rccccc}
\textbf{crystal thickness $\bm{[\mathrm{mm}]}$} & 1 & 2 & 3 & 4 & 5 \\
\textbf{$\bm{N_\gamma}$} & \multicolumn{1}{l}{} & \multicolumn{1}{l}{} & \multicolumn{1}{l}{} & \multicolumn{1}{l}{} & \multicolumn{1}{l}{} \\
$<100~\mathrm{MeV}$, amorphous & 1.1 & 2.6 & 4.6 & 7.4 & 10.9 \\
$<100~\mathrm{MeV}$, $\langle 111 \rangle$ axis & 6.1 & 11.3 & 17.2 & 24.0 & 31.8 \\
full spectrum,  amorphous & 2.3 & 4.7 & 7.5 & 11.0 & 15.1 \\
full spectrum, $\langle 111 \rangle$ axis & 11.0 & 17.6 & 24.0 & 31.0 & 38.8
\end{tabular}
\label{tab:photon_spectrum}
\end{table}
As shown in the Table~\ref{tab:photon_spectrum}, the enhancement in photon yield is higher for the low-energy part of the spectrum compared to that of the full spectrum case. This difference amounts for about 14 \% and illustrates a higher ratio of soft photons produced in the oriented crystal. 
Based on these studies, a 2~mm thick crystal has been selected to be used as a radiator for the hybrid positron source. It provides a good photon yield, a moderate value photon divergence and energy deposition in the crystal.

The simulation of the positron production is performed using a Geant4 code. 
The simulated positron source is based on the target converter using a metal target in the conventional scheme or compound target consisting of a crystal which serves as a radiator of photons followed by the amorphous metal target used for positron generation in the hybrid scheme displayed in Figure~\ref{fig:sources_twoStage} up. The distance between the targets in hybrid scheme is kept 0.2~m allowing the installation of a collimator to clean up the halo of photon and charged particle distribution after the crystal.  As the materials with high Z are preferable for the positron converters, tungsten (Z=74) was chosen for the target-converter material in the simulations. 
The incident beam has been set up as a 6~GeV electron beam with the angular divergence 0.1~mrad and the r.m.s. transverse beam size of 0.5~mm. 
After the optimization studies regarding the positron yield at the target exit, a 17.6~mm thick target made of tungsten has been used to simulate the production of the positrons in the conventional scheme and 2~mm tungsten crystal followed by the 10~mm thick amorphous tungsten target in the hybrid scheme. This choice provides the maximum positron production rate for both schemes.
In the real design of positron sources, the final target thicknesses are defined as the trade off between the positron production rate and the deposited power in the target.
The summary of the obtained results is given in Table~\ref{tab:Target_results}. For the comparable values of the positron yield, the deposited energy and the PEDD are lower for the hybrid scheme compared to the conventional one.

        

\begin{table}[]
\centering
\caption{Results for the positron production simulations.}
\begin{tabular}{rcc}
\textbf{scheme} & conventional & hybrid\footnotemark[1] \\
\textbf{target thickness $\bm{[\mathrm{mm}]}$} & 17.6 & 2 $+$ 10 \\
\textbf{$\bm{e^+}$ production rate $\bm{[N_{e^+}/N_{e^-}]}$} & 14.4 & 15.1 \\
\textbf{target deposited energy $\bm{[\mathrm{GeV}/e^-]}$} & 1.44 & 0.946 \\
\textbf{PEDD $\bm{[\mathrm{GeV}/\mathrm{mm}^3/e^-]}$} & 0.0416 & 0.0156
\end{tabular}
\footnotetext[1]{The values are given for the amorphous target-converter installed after the crystal target.}
\label{tab:Target_results}
\end{table}

The PEDD with 2 bunches of electrons (3.5~nC/bunch) is expected to be 14.5~J/g and 5.5~J/g in the case of conventional and hybrid scheme, respectively. With these parameters, also considered for the FCC-ee positron source, the obtained values stay well below the 35~J/g, limit imposed by the SLC target breakdown. The energy deposited in the target and PEDD, thus, are 34\% and 63\%  respectively lower in the case of the hybrid scheme.

\begin{figure}[h]
\centering
\begin{subfigure}{.5\textwidth}
  \centering
  \includegraphics[width=1.\linewidth]{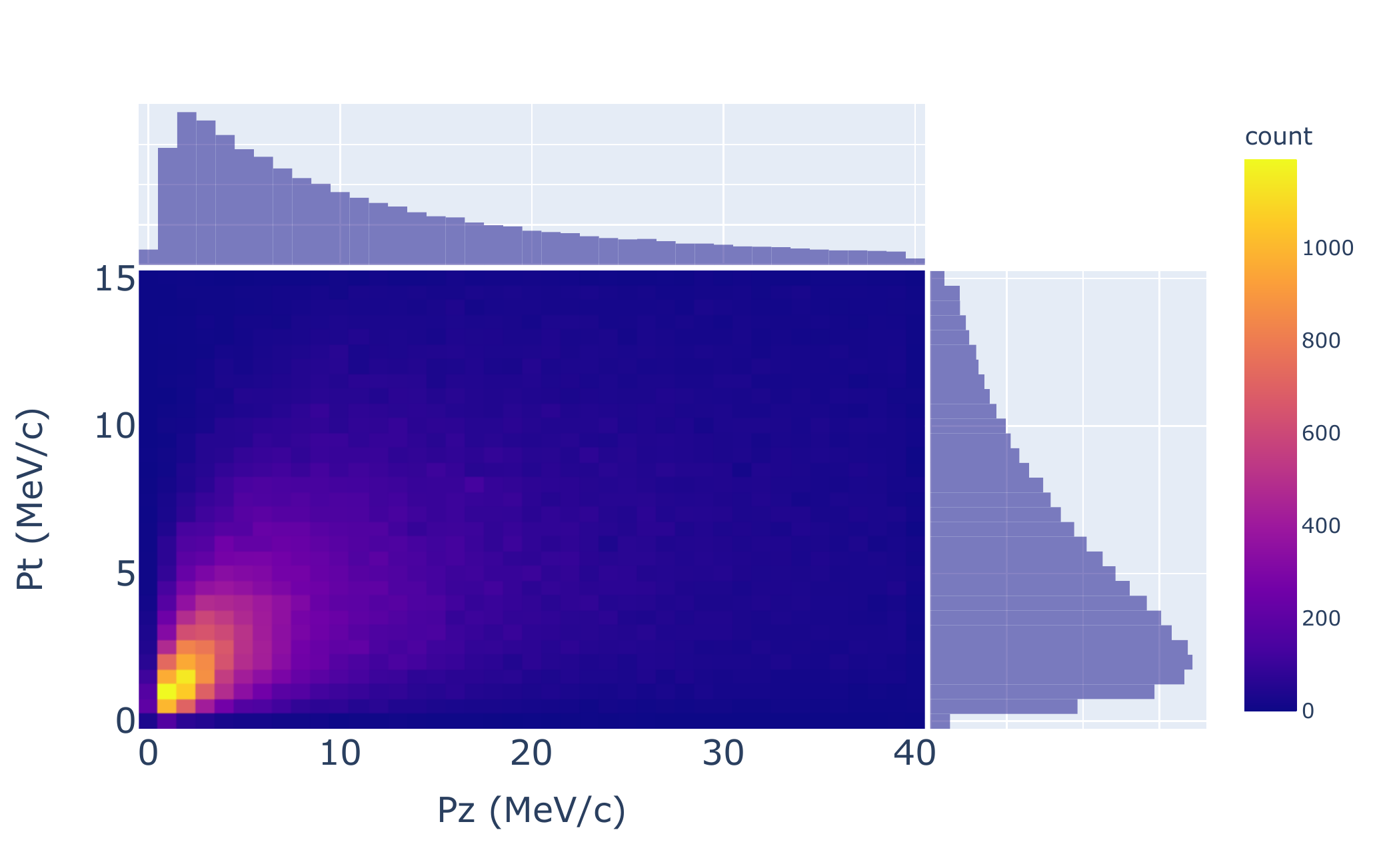}
  \caption{}
  \label{fig:sub1}
\end{subfigure}%
\begin{subfigure}{.5\textwidth}
  \centering
  \includegraphics[width=1.\linewidth]{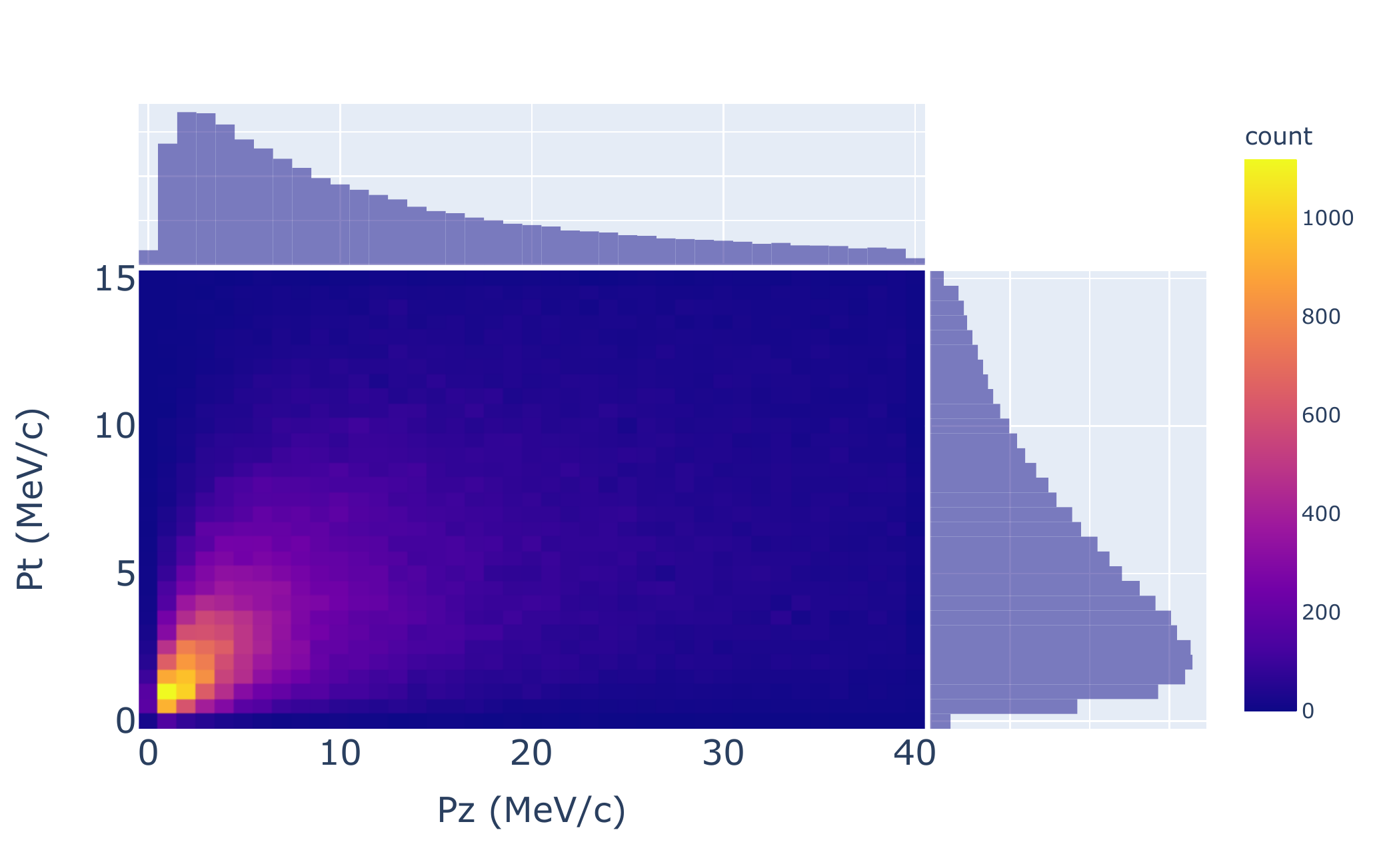}
  \caption{}
  \label{fig:sub2}
\end{subfigure}
\caption{($P_Z$ - $P_t$) Longitudinal-transverse momentum diagram for the positron produced for the conventional scheme (a) and hybrid scheme (b).}
\label{fig:Positrons_pt_pl}
\end{figure}

The created positrons are captured at the target exit by a magnetic matching device being part of the positron capture system. The chosen one is the Adiabatic Matching Device (AMD)~\cite{Chehab:197428}, which exhibits a large momentum acceptance. The generated positrons can be  observed in a ($P_Z$ - $P_t$ ) diagram, as any matching device can be characterized by its acceptance in longitudinal ($P_Z$) and transverse ($P_t$) momenta. Figure~\ref{fig:Positrons_pt_pl} shows such diagrams for the conventional and hybrid schemes. The highest density for the created positrons are in some restricted area of the domains. It allows choosing the parameters of the capture system in order to collect a large number of positrons. 

\begin{figure}[!ht]
\centering
\begin{subfigure}{.5\textwidth}
  \centering
  \includegraphics[width=1.\linewidth]{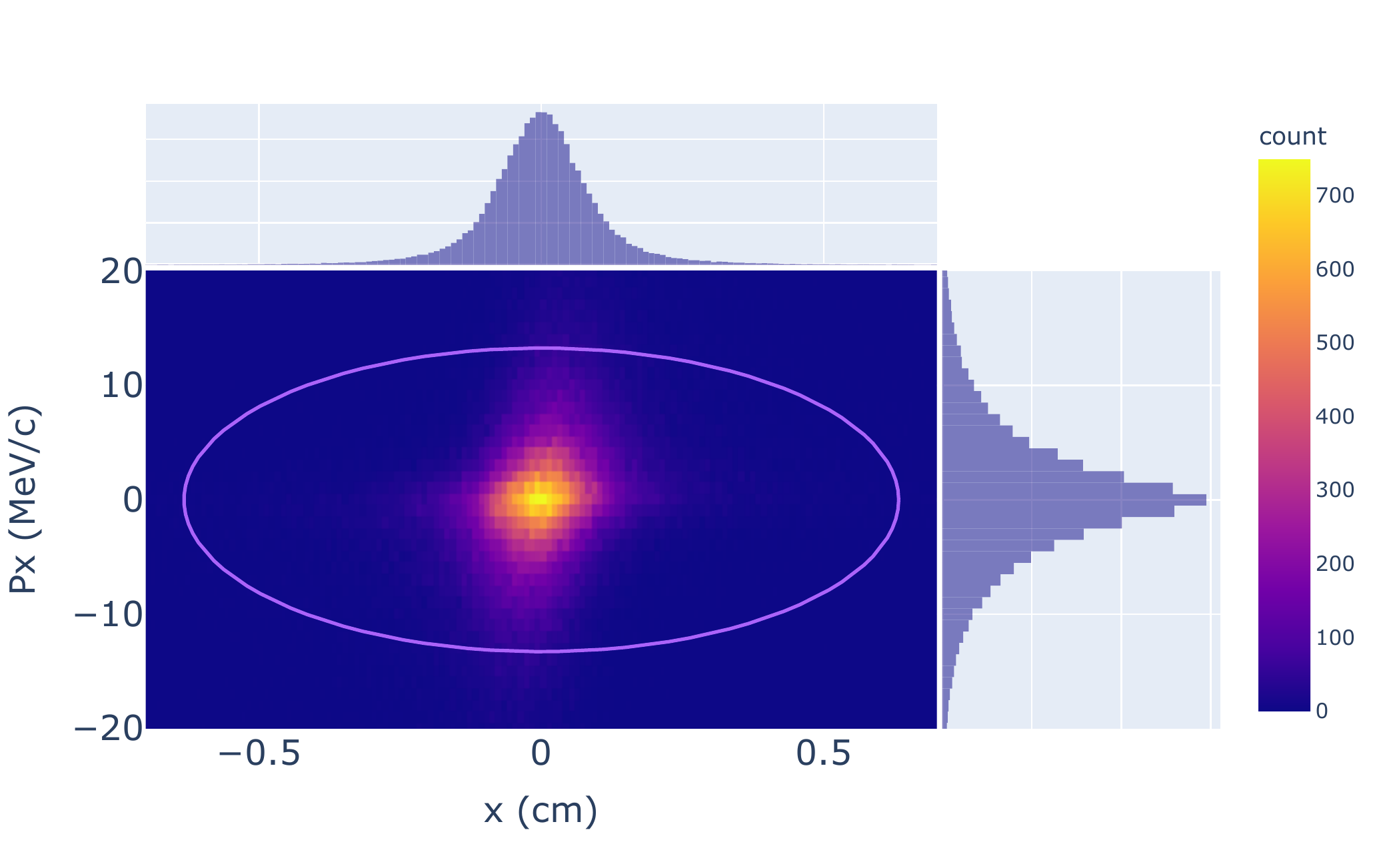}
  \caption{}
  \label{fig:sub1}
\end{subfigure}%
\begin{subfigure}{.5\textwidth}
  \centering
  \includegraphics[width=1.\linewidth]{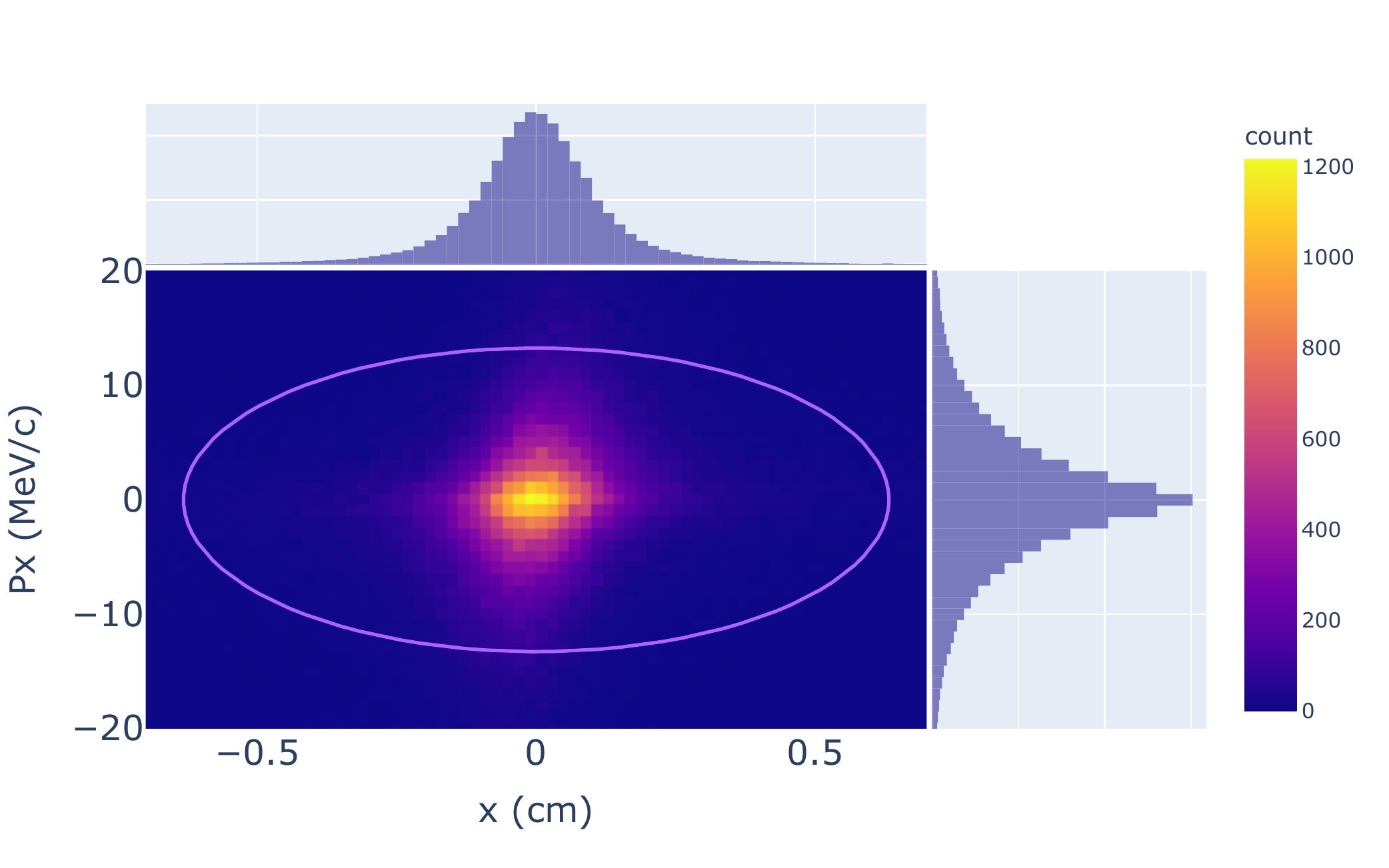}
  \caption{}
  \label{fig:sub2}
\end{subfigure}
\caption{Positron transverse phase space at the target exit and the 
acceptance boundary for the conventional scheme (a) and hybrid scheme (b). About 80\% of the generated positrons are within the theoretical transverse acceptance in x - P$_x$ plane for both schemes.}
\label{fig:Positrons_px_x}
\end{figure}

The positron phase space at the target exit, presented in Figure~\ref{fig:Positrons_px_x} for both conventional and hybrid schemes, is useful for determining the acceptance parameters of the matching system as the geometrical acceptance (maximum horizontal/vertical dimension) and the maximum (horizontal/vertical) transverse momentum.
According to the theoretical considerations for this kind of the focusing system~\cite{Chehab:197428}, the acceptance ellipses in the phase space are calculated and shown in Figure~\ref{fig:Positrons_px_x}, In such a way

\begin{align}
r_0^{max} =\sqrt{ \frac{B_s}{B_0}}a && P_x^{max} = e\sqrt{B_s B_0}a,
\label{eq:AMDacc}
\end{align}
where $B_0$ and $B_s$ are the maximum and minimum values, respectively, of the AMD magnetic field and $a$ is the aperture radius in the capture section. 
The maximum transverse momentum $P_x^{max}$ corresponds to the configuration, when the target is placed in the focusing field of the AMD, while it is twice smaller if the target is placed in a zero-field region.
For the chosen parameters of the AMD ($B_0$ = 7~Tesla, $B_s$ = 0.7~ Tesla and $a$ = 20~mm), we get $r_0^{max}$ to be 6.3~mm  and $P_x^{max}$ equals to 13.3~MeV/$c$.
Taking into account the longitudinal acceptance, which is $P_Z\leq$~34.9~MeV/$c$ with the current parameters, the capture efficiency (part of the transverse and longitudinal accepted phase space) represents about 60\% for both schemes (conventional and hybrid).
In such a way, given the same estimated value of the capture efficiency, the advantages of the hybrid scheme are well highlighted in Table~\ref{tab:Target_results} being lower values of the target deposited energy the PEDD.

\section{Conclusions and outlook}
\label{section:conclusion}

The challenge of designing and building future lepton colliders requires to study in depth new solutions for positron sources. 
Positron sources are in fact complex devices, where each stage (production, capture, acceleration, and injection strategy) has an impact on the final efficiency of the system. Therefore, a start-to-end simulations and optimization have to be performed to evaluate the accepted positron yield, which will be delivered to the interaction point of main accelerator.
 In this paper we presented new results for the study of the radiation emitted by a $W$ crystal in the condition of axial alignment with an impinging 5.6~GeV electron beam. This leads to the validation of a detailed simulation tool for radiation emission in channeling orientation.
 
Two schemes using a conventional and a hybrid target respectively are being considered in this paper implementing the channeling radiation simulation along with a Geant4 simulation. Our studies show that both schemes provide a comparable positron yield at production and capture efficiency. However, as far as reliability of the target is concerned, the hybrid scheme is more attractive allowing lower deposited power and PEDD in the production target. This results are of great interest for the positron source of future circular colliders, such as FCC-ee or CepC. Nevertheless, a detailed analysis of thermal load in the target including peak stress and fatigue caused by the heat cycling load from the pulsed operation are indispensable and need to be further investigated to make a real proposal for a future high-intensity positron source.

\section{Acknowledgements}
We gratefully acknowledge the beamtime provided by the test beam facility at DESY Hamburg (Germany), a member of the Helmholtz Association (HGF). This work was partially supported by INFN CSN5 STORM experiment and the European Commission through the H2020-MSCA-RISE N-LIGHT project (GA. 872196). A. Sytov acknowledges support by the European Commission through the H2020-MSCA-IF TRILLION project (GA. 101032975). We acknowledge the CINECA award under the ISCRA initiative for the availability of high performance computing resources and support. Thu Nhi Tran Caliste is thankfully acknowledged for technical assistance at beamline BM05 of ESRF. We also thank the LNF-BTF and the PADME collaboration, expecially Luca Foggetta, for giving us the BGO calorimeter, and the LNF-SPCM, where Tommaso Napolitano and Fabrizio Angeloni provided the mechanical support structure.
I.~Chaikovska wishes to acknowledge the support from ANR (Agence Nationale de la Recherche) under Grant No: ANR-21-CE31-0007 and the European Union’s Horizon 2020 Research and Innovation programme under Grant Agreement No 101004730.


\bibliography{bibliography}


\begin{thebibliography}{66}
\ifx \bisbn   \undefined \def \bisbn  #1{ISBN #1}\fi
\ifx \binits  \undefined \def \binits#1{#1}\fi
\ifx \bauthor  \undefined \def \bauthor#1{#1}\fi
\ifx \batitle  \undefined \def \batitle#1{#1}\fi
\ifx \bjtitle  \undefined \def \bjtitle#1{#1}\fi
\ifx \bvolume  \undefined \def \bvolume#1{\textbf{#1}}\fi
\ifx \byear  \undefined \def \byear#1{#1}\fi
\ifx \bissue  \undefined \def \bissue#1{#1}\fi
\ifx \bfpage  \undefined \def \bfpage#1{#1}\fi
\ifx \blpage  \undefined \def \blpage #1{#1}\fi
\ifx \burl  \undefined \def \burl#1{\textsf{#1}}\fi
\ifx \doiurl  \undefined \def \doiurl#1{\url{https://doi.org/#1}}\fi
\ifx \betal  \undefined \def \betal{\textit{et al.}}\fi
\ifx \binstitute  \undefined \def \binstitute#1{#1}\fi
\ifx \binstitutionaled  \undefined \def \binstitutionaled#1{#1}\fi
\ifx \bctitle  \undefined \def \bctitle#1{#1}\fi
\ifx \beditor  \undefined \def \beditor#1{#1}\fi
\ifx \bpublisher  \undefined \def \bpublisher#1{#1}\fi
\ifx \bbtitle  \undefined \def \bbtitle#1{#1}\fi
\ifx \bedition  \undefined \def \bedition#1{#1}\fi
\ifx \bseriesno  \undefined \def \bseriesno#1{#1}\fi
\ifx \blocation  \undefined \def \blocation#1{#1}\fi
\ifx \bsertitle  \undefined \def \bsertitle#1{#1}\fi
\ifx \bsnm \undefined \def \bsnm#1{#1}\fi
\ifx \bsuffix \undefined \def \bsuffix#1{#1}\fi
\ifx \bparticle \undefined \def \bparticle#1{#1}\fi
\ifx \barticle \undefined \def \barticle#1{#1}\fi
\bibcommenthead
\ifx \bconfdate \undefined \def \bconfdate #1{#1}\fi
\ifx \botherref \undefined \def \botherref #1{#1}\fi
\ifx \url \undefined \def \url#1{\textsf{#1}}\fi
\ifx \bchapter \undefined \def \bchapter#1{#1}\fi
\ifx \bbook \undefined \def \bbook#1{#1}\fi
\ifx \bcomment \undefined \def \bcomment#1{#1}\fi
\ifx \oauthor \undefined \def \oauthor#1{#1}\fi
\ifx \citeauthoryear \undefined \def \citeauthoryear#1{#1}\fi
\ifx \endbibitem  \undefined \def \endbibitem {}\fi
\ifx \bconflocation  \undefined \def \bconflocation#1{#1}\fi
\ifx \arxivurl  \undefined \def \arxivurl#1{\textsf{#1}}\fi
\csname PreBibitemsHook\endcsname

\bibitem{FCC_CDR_2}
\begin{barticle}
\bauthor{\bsnm{Abada}, \binits{A.e.a.}}:
\batitle{Fcc-ee: The lepton collider}.
\bjtitle{The European Physical Journal Special Topics}
\bvolume{228}(\bissue{2}),
\bfpage{261}--\blpage{623}
(\byear{2019})
\end{barticle}
\endbibitem

\bibitem{Chaikovska:2019ztn}
\begin{bchapter}
\bauthor{\bsnm{Chaikovska}, \binits{I.}}, \betal:
\bctitle{{Positron source for FCC-ee}}.
In: \bbtitle{{10th International Particle Accelerator Conference}},
p. \bfpage{003}
(\byear{2019}).
\doiurl{10.18429/JACoW-IPAC2019-MOPMP003}
\end{bchapter}
\endbibitem

\bibitem{Alesini:2019tlf}
\begin{botherref}
\oauthor{\bsnm{Alesini}, \binits{D.}}, et al.:
{Positron driven muon source for a muon collider}
(2019)
{\href{https://arxiv.org/abs/1905.05747}{{arXiv:1905.05747}}}
{[physics.acc-ph]}
\end{botherref}
\endbibitem

\bibitem{Amapane:2019oog}
\begin{barticle}
\bauthor{\bsnm{Amapane}, \binits{N.}}, \betal:
\batitle{{Study of muon pair production from positron annihilation at threshold
  energy}}.
\bjtitle{JINST}
\bvolume{15}(\bissue{01}),
\bfpage{01036}
(\byear{2020})
{\href{https://arxiv.org/abs/1909.13716}{{arXiv:1909.13716}}}
{[hep-ex]}.
\doiurl{10.1088/1748-0221/15/01/P01036}
\end{barticle}
\endbibitem

\bibitem{Amapane:2021npr}
\begin{botherref}
\oauthor{\bsnm{Amapane}, \binits{N.}}, et al.:
{Muon detection in electron-positron annihilation for muon collider studies}
(2021)
{\href{https://arxiv.org/abs/2105.12624}{{arXiv:2105.12624}}}
{[physics.ins-det]}
\end{botherref}
\endbibitem

\bibitem{cesarini:2021}
\begin{barticle}
\bauthor{\bsnm{Cesarini}, \binits{G.}},
\bauthor{\bsnm{Antonelli}, \binits{M.}},
\bauthor{\bsnm{Anulli}, \binits{F.}},
\bauthor{\bsnm{Bauce}, \binits{M.}},
\bauthor{\bsnm{Biagini}, \binits{M.E.}},
\bauthor{\bsnm{Blanco-Garc{\'\i}a}, \binits{O.R.}},
\bauthor{\bsnm{Boscolo}, \binits{M.}},
\bauthor{\bsnm{Casaburo}, \binits{F.}},
\bauthor{\bsnm{Cavoto}, \binits{G.}},
\bauthor{\bsnm{Ciarma}, \binits{A.}},
\bauthor{\bsnm{Collamati}, \binits{F.}},
\bauthor{\bsnm{Daout}, \binits{C.}},
\bauthor{\bsnm{Li~Voti}, \binits{R.}},
\bauthor{\bsnm{Variola}, \binits{A.}}:
\batitle{Theoretical modeling for the thermal stability of solid targets in a
  positron-driven muon collider}.
\bjtitle{International Journal of Thermophysics}
\bvolume{42}(\bissue{12}),
\bfpage{163}
(\byear{2021}).
\doiurl{10.1007/s10765-021-02913-x}
\end{barticle}
\endbibitem

\bibitem{sheppard2003conventional}
\begin{botherref}
\oauthor{\bsnm{Sheppard}, \binits{J.C.}}:
Conventional positron target for a tesla formatted beam.
LCC-0133,(SLACTN-03-072),(November, 2003)
(2003)
\end{botherref}
\endbibitem

\bibitem{STRAKHOVENKO2005320}
\begin{barticle}
\bauthor{\bsnm{Strakhovenko}, \binits{V.}},
\bauthor{\bsnm{Artru}, \binits{X.}},
\bauthor{\bsnm{Chehab}, \binits{R.}},
\bauthor{\bsnm{Chevallier}, \binits{M.}}:
\batitle{Generation of circularly polarized photons for a linear collider
  polarized positron source}.
\bjtitle{Nuclear Instruments and Methods in Physics Research Section A:
  Accelerators, Spectrometers, Detectors and Associated Equipment}
\bvolume{547}(\bissue{2}),
\bfpage{320}--\blpage{333}
(\byear{2005}).
\doiurl{10.1016/j.nima.2005.03.168}
\end{barticle}
\endbibitem

\bibitem{chaikovska2012polarized}
\begin{botherref}
\oauthor{\bsnm{Chaikovska}, \binits{I.}}:
Polarized positron sources for the future linear colliders.
PhD thesis,
Universit{\'e} Paris Sud-Paris XI
(2012)
\end{botherref}
\endbibitem

\bibitem{balewski1991status}
\begin{botherref}
\oauthor{\bsnm{Balewski}, \binits{K.}},
\oauthor{\bsnm{Bieler}, \binits{M.}},
\oauthor{\bsnm{Bothe}, \binits{W.}},
\oauthor{\bsnm{Bredehoeft}, \binits{K.}},
\oauthor{\bsnm{Brinkmann}, \binits{R.}},
\oauthor{\bsnm{Choroba}, \binits{S.}},
\oauthor{\bsnm{Dwersteg}, \binits{B.}},
\oauthor{\bsnm{Ebert}, \binits{M.}},
\oauthor{\bsnm{Febel}, \binits{A.}},
\oauthor{\bsnm{Fischer}, \binits{R.}}, et al.:
Status report of a 500 gev s-band linear collider study.
Technical report,
Deutsches Elektronen-Synchrotron (DESY)
(1991)
\end{botherref}
\endbibitem

\bibitem{holtkamp1995status}
\begin{bchapter}
\bauthor{\bsnm{Holtkamp}, \binits{N.}}:
\bctitle{The status of the s-band linear collider study}.
In: \bbtitle{Proceedings Particle Accelerator Conference},
vol. \bseriesno{1},
pp. \bfpage{683}--\blpage{685}
(\byear{1995}).
\bcomment{IEEE}
\end{bchapter}
\endbibitem

\bibitem{KUMAKHOV197617}
\begin{barticle}
\bauthor{\bsnm{Kumakhov}, \binits{M.A.}}:
\batitle{On the theory of electromagnetic radiation of charged particles in a
  crystal}.
\bjtitle{Physics Letters A}
\bvolume{57}(\bissue{1}),
\bfpage{17}--\blpage{18}
(\byear{1976}).
\doiurl{10.1016/0375-9601(76)90438-2}
\end{barticle}
\endbibitem

\bibitem{Baier_Katkov_Strakhovenko}
\begin{barticle}
\bauthor{\bsnm{Baier}, \binits{V.N.}},
\bauthor{\bsnm{Katkov}, \binits{V.M.}},
\bauthor{\bsnm{Strakhovenko}, \binits{V.M.}}:
\batitle{Radiation yield of high-energy electrons in thick crystals}.
\bjtitle{physica status solidi (b)}
\bvolume{133}(\bissue{2}),
\bfpage{583}--\blpage{592}
(\byear{1986})
{\href{https://arxiv.org/abs/https://onlinelibrary.wiley.com/doi/pdf/10.1002/pssb.2221330219}{{https://onlinelibrary.wiley.com/doi/pdf/10.1002/pssb.2221330219}}}.
\doiurl{10.1002/pssb.2221330219}
\end{barticle}
\endbibitem

\bibitem{PhysRevLett.53.2371}
\begin{barticle}
\bauthor{\bsnm{Belkacem}, \binits{A.}},
\bauthor{\bsnm{Bologna}, \binits{G.}},
\bauthor{\bsnm{Chevallier}, \binits{M.}},
\bauthor{\bsnm{Clouvas}, \binits{A.}},
\bauthor{\bsnm{Cue}, \binits{N.}},
\bauthor{\bsnm{Gaillard}, \binits{M.J.}},
\bauthor{\bsnm{Genre}, \binits{R.}},
\bauthor{\bsnm{Kimball}, \binits{C.C.}},
\bauthor{\bsnm{Kirsch}, \binits{R.}},
\bauthor{\bsnm{Marsh}, \binits{B.}},
\bauthor{\bsnm{Peigneux}, \binits{J.P.}},
\bauthor{\bsnm{Poizat}, \binits{J.C.}},
\bauthor{\bsnm{Remillieux}, \binits{J.}},
\bauthor{\bsnm{Sillou}, \binits{D.}},
\bauthor{\bsnm{Spighel}, \binits{M.}},
\bauthor{\bsnm{Sun}, \binits{C.R.}}:
\batitle{Observation of enhanced pair creation for 50-110-gev photons in an
  aligned ge crystal}.
\bjtitle{Phys. Rev. Lett.}
\bvolume{53},
\bfpage{2371}--\blpage{2373}
(\byear{1984}).
\doiurl{10.1103/PhysRevLett.53.2371}
\end{barticle}
\endbibitem

\bibitem{BAIER1995147}
\begin{barticle}
\bauthor{\bsnm{Baier}, \binits{V.N.}},
\bauthor{\bsnm{Katkov}, \binits{V.M.}},
\bauthor{\bsnm{Strakhovenko}, \binits{V.M.}}:
\batitle{Electromagnetic showers in crystals at gev energies}.
\bjtitle{Nuclear Instruments and Methods in Physics Research Section B: Beam
  Interactions with Materials and Atoms}
\bvolume{103}(\bissue{2}),
\bfpage{147}--\blpage{155}
(\byear{1995}).
\doiurl{10.1016/0168-583X(95)00653-2}
\end{barticle}
\endbibitem

\bibitem{BAIER1999403}
\begin{barticle}
\bauthor{\bsnm{Baier}, \binits{V.N.}},
\bauthor{\bsnm{Strakhovenko}, \binits{V.M.}}:
\batitle{Crystal assisted positron source efficiency in multi-gev energy
  range}.
\bjtitle{Nuclear Instruments and Methods in Physics Research Section B: Beam
  Interactions with Materials and Atoms}
\bvolume{155}(\bissue{4}),
\bfpage{403}--\blpage{408}
(\byear{1999}).
\doiurl{10.1016/S0168-583X(99)00254-2}
\end{barticle}
\endbibitem

\bibitem{Bandiera18}
\begin{barticle}
\bauthor{\bsnm{Bandiera}, \binits{L.}}, \betal:
\batitle{{Strong reduction of the effective radiation length in an axially
  oriented scintillator crystal}}.
\bjtitle{Phys. Rev. Lett.}
\bvolume{121}(\bissue{2}),
\bfpage{021603}
(\byear{2018}).
\doiurl{10.1103/physrevlett.121.021603}
\end{barticle}
\endbibitem

\bibitem{ARTRU1994443}
\begin{barticle}
\bauthor{\bsnm{Artru}, \binits{X.}},
\bauthor{\bsnm{Baier}, \binits{V.N.}},
\bauthor{\bsnm{Chehab}, \binits{R.}},
\bauthor{\bsnm{Jejcic}, \binits{A.}}:
\batitle{Positron source using channeling in a tungsten crystal}.
\bjtitle{Nuclear Instruments and Methods in Physics Research Section A:
  Accelerators, Spectrometers, Detectors and Associated Equipment}
\bvolume{344}(\bissue{3}),
\bfpage{443}--\blpage{454}
(\byear{1994}).
\doiurl{10.1016/0168-9002(94)90865-6}
\end{barticle}
\endbibitem

\bibitem{Chehab:197428}
\begin{botherref}
\oauthor{\bsnm{Chehab}, \binits{R.}}:
{Positron sources},
30
(1989).
\doiurl{10.5170/CERN-1989-005.105}
\end{botherref}
\endbibitem

\bibitem{chehab1989study}
\begin{bchapter}
\bauthor{\bsnm{Chehab}, \binits{R.}},
\bauthor{\bsnm{Couchot}, \binits{F.}},
\bauthor{\bsnm{Nyaiesh}, \binits{A.}},
\bauthor{\bsnm{Richard}, \binits{F.}},
\bauthor{\bsnm{Artru}, \binits{X.}}:
\bctitle{Study of a positron source generated by photons from ultrarelativistic
  channeled particles}.
In: \bbtitle{Proceedings of the 1989 IEEE Particle Accelerator
  Conference,.'Accelerator Science and Technology},
pp. \bfpage{283}--\blpage{285}
(\byear{1989}).
\bcomment{IEEE}
\end{bchapter}
\endbibitem

\bibitem{ARTRU1996246}
\begin{barticle}
\bauthor{\bsnm{Artru}, \binits{X.}},
\bauthor{\bsnm{Baier}, \binits{V.N.}},
\bauthor{\bsnm{Baier}, \binits{T.V.}}, \betal:
\batitle{Axial channeling of relativistic electrons in crystals as a source for
  positron production}.
\bjtitle{Nuclear Instruments and Methods in Physics Research Section B: Beam
  Interactions with Materials and Atoms}
\bvolume{119}(\bissue{1}),
\bfpage{246}--\blpage{252}
(\byear{1996}).
\doiurl{10.1016/0168-583X(96)00320-5}
\end{barticle}
\endbibitem

\bibitem{baier1968processes}
\begin{barticle}
\bauthor{\bsnm{Baier}, \binits{V.}},
\bauthor{\bsnm{Katkov}, \binits{V.}}:
\batitle{Processes involved in the motion of high energy particles in a
  magnetic field}.
\bjtitle{Sov. Phys. JETP}
\bvolume{26},
\bfpage{854}
(\byear{1968})
\end{barticle}
\endbibitem

\bibitem{Baier_Katkov_Strakhovenko0}
\begin{botherref}
\oauthor{\bsnm{Baier}, \binits{V.N.}},
\oauthor{\bsnm{Katkov}, \binits{V.M.}},
\oauthor{\bsnm{Strakhovenko}, \binits{V.M.}}:
Electromagnetic processes at high energies in oriented single crystals.
(World Scientific, Singapore, 1998)
\end{botherref}
\endbibitem

\bibitem{XAVIER1990278}
\begin{barticle}
\bauthor{\bsnm{Artru}, \binits{X.}}:
\batitle{A simulation code for channeling radiation by ultrarelativistic
  electrons or positrons}.
\bjtitle{Nuclear Instruments and Methods in Physics Research Section B: Beam
  Interactions with Materials and Atoms}
\bvolume{48}(\bissue{1}),
\bfpage{278}--\blpage{282}
(\byear{1990}).
\doiurl{10.1016/0168-583X(90)90122-B}
\end{barticle}
\endbibitem

\bibitem{Tikhomirov01}
\begin{barticle}
\bauthor{\bsnm{Tikhomirov}, \binits{V.V.}}:
\batitle{{Polarization effects accompanying penetration of high-energy
  electrons, positrons and gamma-quanta through crystals}}.
\bjtitle{Rad. effects. and defects in solids}
\bvolume{117}(\bissue{1}),
\bfpage{27}
(\byear{1991}).
\doiurl{10.1080/10420159108220589}
\end{barticle}
\endbibitem

\bibitem{Tikhomirov02}
\begin{barticle}
\bauthor{\bsnm{Tikhomirov}, \binits{V.V.}}:
\batitle{{A new method to calculate the characteristics of radiation and pair
  production under high energies and arbitrary angles of particle incidence
  relative to the crystal planes}}.
\bjtitle{Nucl. Instrum. Methods Phys. Res., Sect. B}
\bvolume{82},
\bfpage{409}
(\byear{1993}).
\doiurl{10.1016/0168-583X(93)95989-I}
\end{barticle}
\endbibitem

\bibitem{Tikhomirov03}
\begin{barticle}
\bauthor{\bsnm{Tikhomirov}, \binits{V.V.}}:
\batitle{{Possibility of observing radiative self-polarization and the
  production of polarized e+e- pairs in crystal at accessible energies}}.
\bjtitle{JETP Lett.}
\bvolume{58}(\bissue{3}),
\bfpage{166}
(\byear{1993})
\end{barticle}
\endbibitem

\bibitem{Tikhomirov04}
\begin{barticle}
\bauthor{\bsnm{Tikhomirov}, \binits{V.V.}}:
\batitle{{To the possibility to observe positron magnetic moment variation
  under the propagation through crystals}}.
\bjtitle{Sov. Yad. Phys.}
\bvolume{57},
\bfpage{2302}
(\byear{1994})
\end{barticle}
\endbibitem

\bibitem{Bandiera19}
\begin{barticle}
\bauthor{\bsnm{Guidi}, \binits{V.}},
\bauthor{\bsnm{Bandiera}, \binits{L.}},
\bauthor{\bsnm{Tikhomirov}, \binits{V.}}:
\batitle{{Radiation generated by single and multiple volume reflection of
  ultrarelativistic electrons and positrons in bent crystals}}.
\bjtitle{Phys. Rev. A}
\bvolume{86},
\bfpage{042903}
(\byear{2012}).
\doiurl{10.1103/PhysRevA.86.042903}
\end{barticle}
\endbibitem

\bibitem{Bandiera21}
\begin{barticle}
\bauthor{\bsnm{Bandiera}, \binits{L.}},
\bauthor{\bsnm{Bagli}, \binits{E.}},
\bauthor{\bsnm{Guidi}, \binits{V.}},
\bauthor{\bsnm{Mazzolari}, \binits{A.}},
\bauthor{\bsnm{Berra}, \binits{A.}},
\bauthor{\bsnm{Lietti}, \binits{D.}},
\bauthor{\bsnm{Prest}, \binits{M.}},
\bauthor{\bsnm{Vallazza}, \binits{E.}},
\bauthor{\bsnm{De~Salvador}, \binits{D.}},
\bauthor{\bsnm{Tikhomirov}, \binits{V.}}:
\batitle{{Broad and intense radiation accompanying multiple volume reflection
  of ultrarelativistic electrons in a bent crystal}}.
\bjtitle{Phys. Rev. Lett.}
\bvolume{111},
\bfpage{255502}
(\byear{2013}).
\doiurl{10.1103/PhysRevLett.111.255502}
\end{barticle}
\endbibitem

\bibitem{Bandiera22}
\begin{barticle}
\bauthor{\bsnm{Bandiera}, \binits{L.}},
\bauthor{\bsnm{Bagli}, \binits{E.}},
\bauthor{\bsnm{Germogli}, \binits{G.}},
\bauthor{\bsnm{Guidi}, \binits{V.}},
\bauthor{\bsnm{Mazzolari}, \binits{A.}},
\bauthor{\bsnm{Backe}, \binits{H.}},
\bauthor{\bsnm{Lauth}, \binits{W.}},
\bauthor{\bsnm{Berra}, \binits{A.}},
\bauthor{\bsnm{Lietti}, \binits{D.}},
\bauthor{\bsnm{Prest}, \binits{M.}},
\bauthor{\bsnm{De~Salvador}, \binits{D.}},
\bauthor{\bsnm{Vallazza}, \binits{E.}},
\bauthor{\bsnm{Tikhomirov}, \binits{V.}}:
\batitle{{Investigation of the electromagnetic radiation emitted by sub-GeV
  electrons in a bent crystal}}.
\bjtitle{Phys. Rev. Lett.}
\bvolume{115},
\bfpage{025504}
(\byear{2015}).
\doiurl{10.1103/PhysRevLett.115.025504}
\end{barticle}
\endbibitem

\bibitem{Bandiera24}
\begin{barticle}
\bauthor{\bsnm{Bandiera}, \binits{L.}},
\bauthor{\bsnm{Bagli}, \binits{E.}},
\bauthor{\bsnm{Guidi}, \binits{V.}},
\bauthor{\bsnm{Tikhomirov}, \binits{V.}}:
\batitle{{RADCHARM++: A C++ routine to compute the electromagnetic radiation
  generated by relativistic charged particles in crystals and complex
  structures}}.
\bjtitle{Nucl. Instrum. Methods Phys. Res., Sect. B}
\bvolume{355},
\bfpage{44}
(\byear{2015}).
\doiurl{10.1016/j.nimb.2015.03.031}
\end{barticle}
\endbibitem

\bibitem{Bandiera25}
\begin{barticle}
\bauthor{\bsnm{Bandiera}, \binits{L.}},
\bauthor{\bsnm{Sytov}, \binits{A.}},
\bauthor{\bsnm{{De Salvador}}, \binits{D.}},
\bauthor{\bsnm{Mazzolari}, \binits{A.}},
\bauthor{\bsnm{Bagli}, \binits{E.}},
\bauthor{\bsnm{Camattari}, \binits{R.}},
\bauthor{\bsnm{Carturan}, \binits{S.}},
\bauthor{\bsnm{Durighello}, \binits{C.}},
\bauthor{\bsnm{Germogli}, \binits{G.}},
\bauthor{\bsnm{Guidi}, \binits{V.}},
\bauthor{\bsnm{Klag}, \binits{P.}},
\bauthor{\bsnm{Lauth}, \binits{W.}},
\bauthor{\bsnm{Maggioni}, \binits{G.}},
\bauthor{\bsnm{Mascagna}, \binits{V.}},
\bauthor{\bsnm{Prest}, \binits{M.}},
\bauthor{\bsnm{Romagnoni}, \binits{M.}},
\bauthor{\bsnm{Soldani}, \binits{M.}},
\bauthor{\bsnm{Tikhomirov}, \binits{V.V.}},
\bauthor{\bsnm{Vallazza}, \binits{E.}}:
\batitle{{Investigation on radiation generated by sub-GeV electrons in
  ultrashort silicon and germanium bent crystals}}.
\bjtitle{Eur. Phys. J. C}
\bvolume{81},
\bfpage{284}
(\byear{2021}).
\doiurl{10.1140/epjc/s10052-021-09071-2}
\end{barticle}
\endbibitem

\bibitem{Sytov2}
\begin{barticle}
\bauthor{\bsnm{Sytov}, \binits{A.}},
\bauthor{\bsnm{Tikhomirov}, \binits{V.}},
\bauthor{\bsnm{Bandiera}, \binits{L.}}:
\batitle{{Simulation code for modeling of coherent effects of radiation
  generation in oriented crystals}}.
\bjtitle{Phys. Rev. Acc. and Beams}
\bvolume{22},
\bfpage{064601}
(\byear{2019}).
\doiurl{10.1103/PhysRevAccelBeams.22.064601}
\end{barticle}
\endbibitem

\bibitem{Mazzolari1}
\begin{barticle}
\bauthor{\bsnm{Mazzolari}, \binits{A.}},
\bauthor{\bsnm{Bagli}, \binits{E.}},
\bauthor{\bsnm{Bandiera}, \binits{L.}},
\bauthor{\bsnm{Guidi}, \binits{V.}},
\bauthor{\bsnm{Backe}, \binits{H.}},
\bauthor{\bsnm{Lauth}, \binits{W.}},
\bauthor{\bsnm{Tikhomirov}, \binits{V.}},
\bauthor{\bsnm{Berra}, \binits{A.}},
\bauthor{\bsnm{Lietti}, \binits{D.}},
\bauthor{\bsnm{Prest}, \binits{M.}},
\bauthor{\bsnm{Vallazza}, \binits{E.}},
\bauthor{\bsnm{De~Salvador}, \binits{D.}}:
\batitle{{Steering of a sub-GeV electron beam through planar channeling
  enhanced by rechanneling}}.
\bjtitle{Phys. Rev. Lett.}
\bvolume{112},
\bfpage{135503}
(\byear{2014}).
\doiurl{10.1103/PhysRevLett.112.135503}
\end{barticle}
\endbibitem

\bibitem{Wistisen1}
\begin{barticle}
\bauthor{\bsnm{Wistisen}, \binits{T.N.}},
\bauthor{\bsnm{Mikkelsen}, \binits{R.E.}},
\bauthor{\bsnm{Uggerh\o{}j}, \binits{U.I.}},
\bauthor{\bsnm{Wienands}, \binits{U.}},
\bauthor{\bsnm{Markiewicz}, \binits{T.W.}},
\bauthor{\bsnm{Gessner}, \binits{S.}},
\bauthor{\bsnm{Hogan}, \binits{M.J.}},
\bauthor{\bsnm{Noble}, \binits{R.J.}},
\bauthor{\bsnm{Holtzapple}, \binits{R.}},
\bauthor{\bsnm{Tucker}, \binits{S.}},
\bauthor{\bsnm{Guidi}, \binits{V.}},
\bauthor{\bsnm{Mazzolari}, \binits{A.}},
\bauthor{\bsnm{Bagli}, \binits{E.}},
\bauthor{\bsnm{Bandiera}, \binits{L.}},
\bauthor{\bsnm{Sytov}, \binits{A.}}:
\batitle{{Observation of quasichanneling oscillations}}.
\bjtitle{Phys. Rev. Lett.}
\bvolume{119},
\bfpage{024801}
(\byear{2017}).
\doiurl{10.1103/PhysRevLett.119.024801}
\end{barticle}
\endbibitem

\bibitem{Sytov3}
\begin{barticle}
\bauthor{\bsnm{Sytov}, \binits{A.I.}},
\bauthor{\bsnm{Bandiera}, \binits{L.}},
\bauthor{\bsnm{{De Salvador}}, \binits{D.}},
\bauthor{\bsnm{Mazzolari}, \binits{A.}},
\bauthor{\bsnm{Bagli}, \binits{E.}},
\bauthor{\bsnm{Berra}, \binits{A.}},
\bauthor{\bsnm{Carturan}, \binits{S.}},
\bauthor{\bsnm{Durighello}, \binits{C.}},
\bauthor{\bsnm{Germogli}, \binits{G.}},
\bauthor{\bsnm{Guidi}, \binits{V.}},
\bauthor{\bsnm{Klag}, \binits{P.}},
\bauthor{\bsnm{Lauth}, \binits{W.}},
\bauthor{\bsnm{Maggioni}, \binits{G.}},
\bauthor{\bsnm{Prest}, \binits{M.}},
\bauthor{\bsnm{Romagnoni}, \binits{M.}},
\bauthor{\bsnm{Tikhomirov}, \binits{V.V.}},
\bauthor{\bsnm{Vallazza}, \binits{E.}}:
\batitle{{Steering of Sub-GeV electrons by ultrashort Si and Ge bent
  crystals}}.
\bjtitle{Eur. Phys. J. C}
\bvolume{77},
\bfpage{901}
(\byear{2017}).
\doiurl{10.1140/epjc/s10052-017-5456-7}
\end{barticle}
\endbibitem

\bibitem{CHEHAB200241}
\begin{barticle}
\bauthor{\bsnm{Chehab}, \binits{R.}},
\bauthor{\bsnm{Cizeron}, \binits{R.}},
\bauthor{\bsnm{Sylvia}, \binits{C.}}, \betal:
\batitle{Experimental study of a crystal positron source}.
\bjtitle{Physics Letters B}
\bvolume{525}(\bissue{1}),
\bfpage{41}--\blpage{48}
(\byear{2002}).
\doiurl{10.1016/S0370-2693(01)01395-8}
\end{barticle}
\endbibitem

\bibitem{ARTRU2005762}
\begin{barticle}
\bauthor{\bsnm{Artru}, \binits{X.}},
\bauthor{\bsnm{Baier}, \binits{V.}},
\bauthor{\bsnm{Beloborodov}, \binits{K.}}, \betal:
\batitle{Summary of experimental studies, at cern, on a positron source using
  crystal effects}.
\bjtitle{Nuclear Instruments and Methods in Physics Research Section B: Beam
  Interactions with Materials and Atoms}
\bvolume{240}(\bissue{3}),
\bfpage{762}--\blpage{776}
(\byear{2005}).
\doiurl{10.1016/j.nimb.2005.04.134}
\end{barticle}
\endbibitem

\bibitem{ARTRU2003243}
\begin{barticle}
\bauthor{\bsnm{Artru}, \binits{X.}},
\bauthor{\bsnm{Baier}, \binits{V.}},
\bauthor{\bsnm{Beloborodov}, \binits{K.}}, \betal:
\batitle{Experiment with a crystal-assisted positron source using 6 and 10 gev
  electrons}.
\bjtitle{Nuclear Instruments and Methods in Physics Research Section B: Beam
  Interactions with Materials and Atoms}
\bvolume{201}(\bissue{1}),
\bfpage{243}--\blpage{252}
(\byear{2003}).
\doiurl{10.1016/S0168-583X(02)01180-1}
\end{barticle}
\endbibitem

\bibitem{PhysRevE.67.016502}
\begin{barticle}
\bauthor{\bsnm{Suwada}, \binits{T.}},
\bauthor{\bsnm{Anami}, \binits{S.}},
\bauthor{\bsnm{Chehab}, \binits{R.}}, \betal:
\batitle{Measurement of positron production efficiency from a tungsten
  monocrystalline target using 4- and 8-gev electrons}.
\bjtitle{Phys. Rev. E}
\bvolume{67},
\bfpage{016502}
(\byear{2003}).
\doiurl{10.1103/PhysRevE.67.016502}
\end{barticle}
\endbibitem

\bibitem{PhysRevSTAB.10.073501}
\begin{barticle}
\bauthor{\bsnm{Suwada}, \binits{T.}},
\bauthor{\bsnm{Satoh}, \binits{M.}},
\bauthor{\bsnm{Furukawa}, \binits{K.}},
\bauthor{\bsnm{Kamitani}, \binits{T.}}, \betal:
\batitle{First application of a tungsten single-crystal positron source at the
  kek b factory}.
\bjtitle{Phys. Rev. ST Accel. Beams}
\bvolume{10},
\bfpage{073501}
(\byear{2007}).
\doiurl{10.1103/PhysRevSTAB.10.073501}
\end{barticle}
\endbibitem

\bibitem{Artru:1997fy}
\begin{barticle}
\bauthor{\bsnm{Artru}, \binits{X.}},
\bauthor{\bsnm{Baier}, \binits{V.N.}},
\bauthor{\bsnm{Chehab}, \binits{R.}},
\bauthor{\bsnm{Chevallier}, \binits{M.}},
\bauthor{\bsnm{Dubrovin}, \binits{M.S.}},
\bauthor{\bsnm{Jejcic}, \binits{A.}},
\bauthor{\bsnm{Silva}, \binits{J.}}:
\batitle{{Positron sources using channeling: A Comparison with conventional
  targets}}.
\bjtitle{Part. Accel.}
\bvolume{59},
\bfpage{19}--\blpage{41}
(\byear{1998})
\end{barticle}
\endbibitem

\bibitem{Artru:2008zz}
\begin{barticle}
\bauthor{\bsnm{Artru}, \binits{X.}},
\bauthor{\bsnm{Chehab}, \binits{R.}},
\bauthor{\bsnm{Chevallier}, \binits{M.}},
\bauthor{\bsnm{Strakhovenko}, \binits{V.M.}},
\bauthor{\bsnm{Variola}, \binits{A.}},
\bauthor{\bsnm{Vivoli}, \binits{A.}}:
\batitle{{Polarized and unpolarized positron sources for electron-positron
  colliders}}.
\bjtitle{Nucl. Instrum. Meth. B}
\bvolume{266},
\bfpage{3868}--\blpage{3875}
(\byear{2008}).
\doiurl{10.1016/j.nimb.2008.02.086}
\end{barticle}
\endbibitem

\bibitem{Dadoun:1248436}
\begin{botherref}
\oauthor{\bsnm{Dadoun}, \binits{O.}},
\oauthor{\bsnm{Chaikovska}, \binits{I.}},
\oauthor{\bsnm{Chehab}, \binits{R.}},
\oauthor{\bsnm{Poirier}, \binits{F.}},
\oauthor{\bsnm{Rinolfi}, \binits{L.}},
\oauthor{\bsnm{Strakhovenko}, \binits{V.}},
\oauthor{\bsnm{Variola}, \binits{A.}},
\oauthor{\bsnm{Vivoli}, \binits{A.}}:
{Study of an hybrid positron source using channeling for CLIC}.
Technical report,
CERN,
Geneva
(Sep 2009).
\url{https://cds.cern.ch/record/1248436}
\end{botherref}
\endbibitem

\bibitem{chehab2011posipol}
\begin{barticle}
\bauthor{\bsnm{Chehab}, \binits{R.}}:
\batitle{Posipol: from polarized and unpolarized photons to positrons}.
\bjtitle{Il nuovo cimento C}
\bvolume{34}(\bissue{4}),
\bfpage{9}--\blpage{17}
(\byear{2011})
\end{barticle}
\endbibitem

\bibitem{artru2011positron}
\begin{barticle}
\bauthor{\bsnm{Artru}, \binits{X.}},
\bauthor{\bsnm{Chaikovska}, \binits{I.}},
\bauthor{\bsnm{Chehab}, \binits{R.}}, \betal:
\batitle{Positron sources using channeling: a promising device for linear
  colliders}.
\bjtitle{Il nuovo cimento C}
\bvolume{34}(\bissue{4}),
\bfpage{141}--\blpage{148}
(\byear{2011})
\end{barticle}
\endbibitem

\bibitem{SATOH20053}
\begin{barticle}
\bauthor{\bsnm{Satoh}, \binits{M.}},
\bauthor{\bsnm{Suwada}, \binits{T.}},
\bauthor{\bsnm{Furukawa}, \binits{K.}},
\bauthor{\bsnm{Kamitani}, \binits{T.}}, \betal:
\batitle{Experimental study of positron production from silicon and diamond
  crystals by 8-gev channeling electrons}.
\bjtitle{Nuclear Instruments and Methods in Physics Research Section B: Beam
  Interactions with Materials and Atoms}
\bvolume{227}(\bissue{1}),
\bfpage{3}--\blpage{10}
(\byear{2005}).
\doiurl{10.1016/j.nimb.2004.03.088}
\end{barticle}
\endbibitem

\bibitem{SUWADA2006142}
\begin{barticle}
\bauthor{\bsnm{Suwada}, \binits{T.}},
\bauthor{\bsnm{Furukawa}, \binits{K.}},
\bauthor{\bsnm{Kamitani}, \binits{T.}}, \betal:
\batitle{Experimental study of positron production from a 2.55-mm-thick silicon
  crystal target using 8-gev electron beams with high-bunch charges}.
\bjtitle{Nuclear Instruments and Methods in Physics Research Section B: Beam
  Interactions with Materials and Atoms}
\bvolume{252}(\bissue{1}),
\bfpage{142}--\blpage{147}
(\byear{2006}).
\doiurl{10.1016/j.nimb.2006.04.065}
\end{barticle}
\endbibitem

\bibitem{dadoun2012event}
\begin{bchapter}
\bauthor{\bsnm{Dadoun}, \binits{O.}},
\bauthor{\bsnm{Le~Meur}, \binits{G.}},
\bauthor{\bsnm{Touze}, \binits{F.}},
\bauthor{\bsnm{Variola}, \binits{A.}},
\bauthor{\bsnm{Artru}, \binits{X.}},
\bauthor{\bsnm{Chehab}, \binits{R.}},
\bauthor{\bsnm{Chevallier}, \binits{M.}},
\bauthor{\bsnm{Strakhovenko}, \binits{V.}}:
\bctitle{An event generator for crystal source application of the clic positron
  baseline}.
In: \bbtitle{Journal of Physics: Conference Series},
vol. \bseriesno{357},
p. \bfpage{012024}
(\byear{2012}).
\bcomment{IOP Publishing}
\end{bchapter}
\endbibitem

\bibitem{UESUGI201417}
\begin{barticle}
\bauthor{\bsnm{Uesugi}, \binits{Y.}},
\bauthor{\bsnm{Akagi}, \binits{T.}},
\bauthor{\bsnm{Chehab}, \binits{R.}}, \betal:
\batitle{Development of an intense positron source using a crystal-amorphous
  hybrid target for linear colliders}.
\bjtitle{Nuclear Instruments and Methods in Physics Research Section B: Beam
  Interactions with Materials and Atoms}
\bvolume{319},
\bfpage{17}--\blpage{23}
(\byear{2014}).
\doiurl{10.1016/j.nimb.2013.10.025}
\end{barticle}
\endbibitem

\bibitem{ARTRU201560}
\begin{barticle}
\bauthor{\bsnm{Artru}, \binits{X.}},
\bauthor{\bsnm{Chaikovska}, \binits{I.}},
\bauthor{\bsnm{Chehab}, \binits{R.}},
\bauthor{\bsnm{Chevallier}, \binits{M.}}, \betal:
\batitle{Investigations on a hybrid positron source with a granular converter}.
\bjtitle{Nuclear Instruments and Methods in Physics Research Section B: Beam
  Interactions with Materials and Atoms}
\bvolume{355},
\bfpage{60}--\blpage{64}
(\byear{2015}).
\doiurl{10.1016/j.nimb.2015.02.027}
\end{barticle}
\endbibitem

\bibitem{CHAIKOVSKA201758}
\begin{barticle}
\bauthor{\bsnm{Chaikovska}, \binits{I.}},
\bauthor{\bsnm{Chehab}, \binits{R.}},
\bauthor{\bsnm{Guler}, \binits{H.}},
\bauthor{\bsnm{Sievers}, \binits{P.}},
\bauthor{\bsnm{Artru}, \binits{X.}}, \betal:
\batitle{Optimization of an hybrid positron source using channeling}.
\bjtitle{Nuclear Instruments and Methods in Physics Research Section B: Beam
  Interactions with Materials and Atoms}
\bvolume{402},
\bfpage{58}--\blpage{62}
(\byear{2017}).
\doiurl{10.1016/j.nimb.2017.03.012}
\end{barticle}
\endbibitem

\bibitem{maloy2001slc}
\begin{botherref}
\oauthor{\bsnm{Maloy}, \binits{S.}}, et al.:
Slc target analysis.
LANL LA UR-01-1913
\textbf{72}
(2001)
\end{botherref}
\endbibitem

\bibitem{stein2001thermal}
\begin{bchapter}
\bauthor{\bsnm{Stein}, \binits{W.}},
\bauthor{\bsnm{Sunwoo}, \binits{A.}},
\bauthor{\bsnm{Bharadwaj}, \binits{V.}},
\bauthor{\bsnm{Schultz}, \binits{D.}},
\bauthor{\bsnm{Sheppard}, \binits{J.}}:
\bctitle{Thermal shock structural analyses of a positron target}.
In: \bbtitle{PACS2001. Proceedings of the 2001 Particle Accelerator Conference
  (Cat. No. 01CH37268)},
vol. \bseriesno{3},
pp. \bfpage{2111}--\blpage{2113}
(\byear{2001}).
\bcomment{IEEE}
\end{bchapter}
\endbibitem

\bibitem{PhysRevSTAB.6.091003}
\begin{barticle}
\bauthor{\bsnm{Artru}, \binits{X.}},
\bauthor{\bsnm{Chehab}, \binits{R.}},
\bauthor{\bsnm{Chevallier}, \binits{M.}},
\bauthor{\bsnm{Strakhovenko}, \binits{V.}}:
\batitle{Advantages of axially aligned crystals used in positron production at
  future linear colliders}.
\bjtitle{Phys. Rev. ST Accel. Beams}
\bvolume{6},
\bfpage{091003}
(\byear{2003}).
\doiurl{10.1103/PhysRevSTAB.6.091003}
\end{barticle}
\endbibitem

\bibitem{azadegan2013positron}
\begin{barticle}
\bauthor{\bsnm{Azadegan}, \binits{B.}},
\bauthor{\bsnm{Mahdipour}, \binits{A.}},
\bauthor{\bsnm{Dabagov}, \binits{S.}},
\bauthor{\bsnm{Wagner}, \binits{W.}}:
\batitle{Positron energy distributions from a hybrid positron source based on
  channeling radiation}.
\bjtitle{Nuclear Instruments and Methods in Physics Research Section B: Beam
  Interactions with Materials and Atoms}
\bvolume{309},
\bfpage{56}--\blpage{58}
(\byear{2013})
\end{barticle}
\endbibitem

\bibitem{cheng2012positron}
\begin{barticle}
\bauthor{\bsnm{Cheng-Hai}, \binits{X.}},
\bauthor{\bsnm{Chehab}, \binits{R.}},
\bauthor{\bsnm{Sievers}, \binits{P.}},
\bauthor{\bsnm{Artru}, \binits{X.}}, \betal:
\batitle{A positron source using an axially oriented crystal associated to a
  granular amorphous converter}.
\bjtitle{Chinese Physics C}
\bvolume{36}(\bissue{9}),
\bfpage{871}
(\byear{2012})
\end{barticle}
\endbibitem

\bibitem{Chaikovska:IPAC17-WEPIK002}
\begin{bchapter}
\bauthor{\bsnm{Chaikovska}, \binits{I.}}, \betal:
\bctitle{Experimental activities on high intensity positron sources using
  channeling}.
In: \bbtitle{Proc. 8th Int. Particle Accelerator Conf. (IPAC'17)},
pp. \bfpage{2910}--\blpage{2913}
(\byear{2017}).
\doiurl{10.18429/JACoW-IPAC2017-WEPIK002}
\end{bchapter}
\endbibitem

\bibitem{2019_DESY}
\begin{barticle}
\bauthor{\bsnm{Diener}, \binits{R.}}, \betal:
\batitle{{The DESY II test beam facility}}.
\bjtitle{Nuclear Instruments and Methods in Physics Research Section A:
  Accelerators, Spectrometers, Detectors and Associated Equipment}
\bvolume{922},
\bfpage{265}--\blpage{286}
(\byear{2019}).
\doiurl{10.1016/j.nima.2018.11.133}
\end{barticle}
\endbibitem

\bibitem{2013_Lietti}
\begin{barticle}
\bauthor{\bsnm{Lietti}, \binits{D.}},
\bauthor{\bsnm{Berra}, \binits{A.}},
\bauthor{\bsnm{Prest}, \binits{M.}},
\bauthor{\bsnm{Vallazza}, \binits{E.}}:
\batitle{{A microstrip silicon telescope for high performance particle
  tracking}}.
\bjtitle{Nucl. Instrum. Meth. A}
\bvolume{729},
\bfpage{527}--\blpage{536}
(\byear{2013}).
\doiurl{10.1016/j.nima.2013.07.066}
\end{barticle}
\endbibitem

\bibitem{2021_Soldani}
\begin{barticle}
\bauthor{\bsnm{Soldani}, \binits{M.}}, \betal:
\batitle{{Next-generation ultra-compact calorimeters based on oriented
  crystals}}.
\bjtitle{PoS}
\bvolume{ICHEP2020},
\bfpage{872}
(\byear{2021}).
\doiurl{10.22323/1.390.0872}
\end{barticle}
\endbibitem

\bibitem{abbiendi2021study}
\begin{barticle}
\bauthor{\bsnm{Abbiendi}, \binits{G.}}, \betal:
\batitle{{A study of muon-electron elastic scattering in a test beam}}.
\bjtitle{Journal of Instrumentation}
\bvolume{16}(\bissue{06}),
\bfpage{06005}
(\byear{2021}).
\doiurl{10.1088/1748-0221/16/06/p06005}
\end{barticle}
\endbibitem

\bibitem{2003_geant}
\begin{barticle}
\bauthor{\bsnm{Agostinelli}, \binits{S.}}, \betal:
\batitle{{Geant4—a simulation toolkit}}.
\bjtitle{Nuclear Instruments and Methods in Physics Research Section A:
  Accelerators, Spectrometers, Detectors and Associated Equipment}
\bvolume{506}(\bissue{3}),
\bfpage{250}--\blpage{303}
(\byear{2003}).
\doiurl{10.1016/S0168-9002(03)01368-8}
\end{barticle}
\endbibitem

\bibitem{Tikhomirov1}
\begin{botherref}
\oauthor{\bsnm{Tikhomirov}, \binits{V.}}:
{A benchmark construction of positron crystal undulator}
(2015)
{\href{https://arxiv.org/abs/1502.06588}{{arXiv:1502.06588}}}
{[physics.acc-ph]}
\end{botherref}
\endbibitem

\bibitem{Baryshevsky1}
\begin{barticle}
\bauthor{\bsnm{Baryshevsky}, \binits{V.G.}},
\bauthor{\bsnm{Tikhomirov}, \binits{V.}}:
\batitle{{Crystal undulators: from the prediction to the mature simulations}}.
\bjtitle{Nucl. Instrum. Methods Phys. Res., Sect. B}
\bvolume{309},
\bfpage{30}
(\byear{2013}).
\doiurl{10.1016/j.nimb.2013.03.013}
\end{barticle}
\endbibitem

\end{thebibliography}


\end{document}